\documentclass[aps,prl,twocolumn,showpacs]{revtex4}
\usepackage{graphicx}
\usepackage{epstopdf}
\usepackage{dcolumn} 
\usepackage{bm}      
\usepackage{amsmath}
\usepackage{amssymb} 
\usepackage{array}
\usepackage[caption=false]{subfig}
\usepackage[bookmarks=false,linkcolor=blue,urlcolor=blue,colorlinks,citecolor=blue]{hyperref}
\usepackage{exscale,relsize}
\usepackage{dsfont}
\usepackage{bbold}

\DeclareMathOperator{\sgn}{sgn}

\DeclareMathOperator{\Imag}{Im}
\makeatletter

\begin{document}

\title{Signatures of Majorana Zero Modes in Spin-Resolved Current Correlations}
\author{Arbel Haim$^1$, Erez Berg$^1$, Felix von Oppen$^2$ and Yuval Oreg$^1$}
\affiliation{$^1$Department of Condensed Matter Physics$,$ Weizmann Institute of Science$,$ Rehovot$,$ 76100$,$ Israel\\
\mbox{$^2$Dahlem Center for Complex Quantum Systems and Fachbereich Physik, Freie Universit\"at Berlin, 14195 Berlin, Germany}}
\date{\today}

\begin{abstract}
We consider a normal lead coupled to a Majorana bound state. We show that the spin-resolved current correlations exhibit unique features which distinguish Majorana bound states from other low-energy resonances. In particular, the spin-up and spin-down currents from a Majorana bound state are anticorrelated at low bias voltages, and become uncorrelated at higher voltages. This behavior is independent of the exact form of coupling to the lead, and of the direction of the spin polarization. In contrast, an ordinary low-energy Andreev bound state gives rise to a positive correlation between the spin-up and spin-down currents, and this spin-resolved current-current correlation approaches a nonzero constant at high bias voltages. We discuss experimental setups in which this effect can be measured.
\end{abstract}

\maketitle

\emph{Introduction.---}Majorana fermions in condensed matter systems~\cite{Alicea2012,Beenakker2013} are zero-energy excitations which behave as particles which are their own antiparticles. The interest in phases which host such Majorana bound states (MBSs) stems largely from their topological nature; such a state stores information nonlocally. A consequence of this property is that such systems are insensitive to local perturbations and to decoherence~\cite{Kitaev2001}. This, along with their non-Abelian exchange statistics~\cite{ivanov2001non,stern2004geometric,nayak2008non,alicea2011non}, makes them a potential platform for fault-tolerant quantum information processing~\cite{Kitaev2003}.

In recent years various mechanisms have been proposed theoretically for realizing MBSs~\cite{read2000paired,ivanov2001non,moore1991nonabelions,fu2008superconducting,fu2009josephson,Sau2010Generic,Alicea2010Majorana,lutchyn2010majorana,oreg2010helical}. Specifically, it was predicted that a semiconductor nanowire with strong spin-orbit coupling in proximity to an $s$-wave superconductor and subject to a magnetic field can host localized MBSs~\cite{lutchyn2010majorana,oreg2010helical}. This proposal prompted a series of transport experiments ~\cite{mourik2012signatures,Deng2012a,Das2012zero,churchill2013superconductor,Finck2013} reporting the observation of a zero-bias conductance peak (ZBCP), consistent with the presence of a MBS~\cite{Law2009majorana,Bolech2007Observing,Fidkowski2012universal}. Very recently zero-energy states have also been detected at the ends of a ferromagnetic atomic chain deposited on a superconductor~\cite{Nadj-Perge2014observation}. While these experiments are promising, it has been suggested that the ZBCP can also appear in the absence of a MBS as a result of other mechanisms~\cite{Pikulin2012,kells2012near,liu2012zero,sasaki2000kondo,Bagrets2012Class,lee2013spin}.
It is therefore crucial to have a physical signature beyond the ZBCP, which will be able to distinguish the MBS from other possible physical scenarios.

\begin{figure}
\includegraphics[clip=true,trim =0cm 3.2cm 0cm 3.2cm,width=0.4\textwidth]{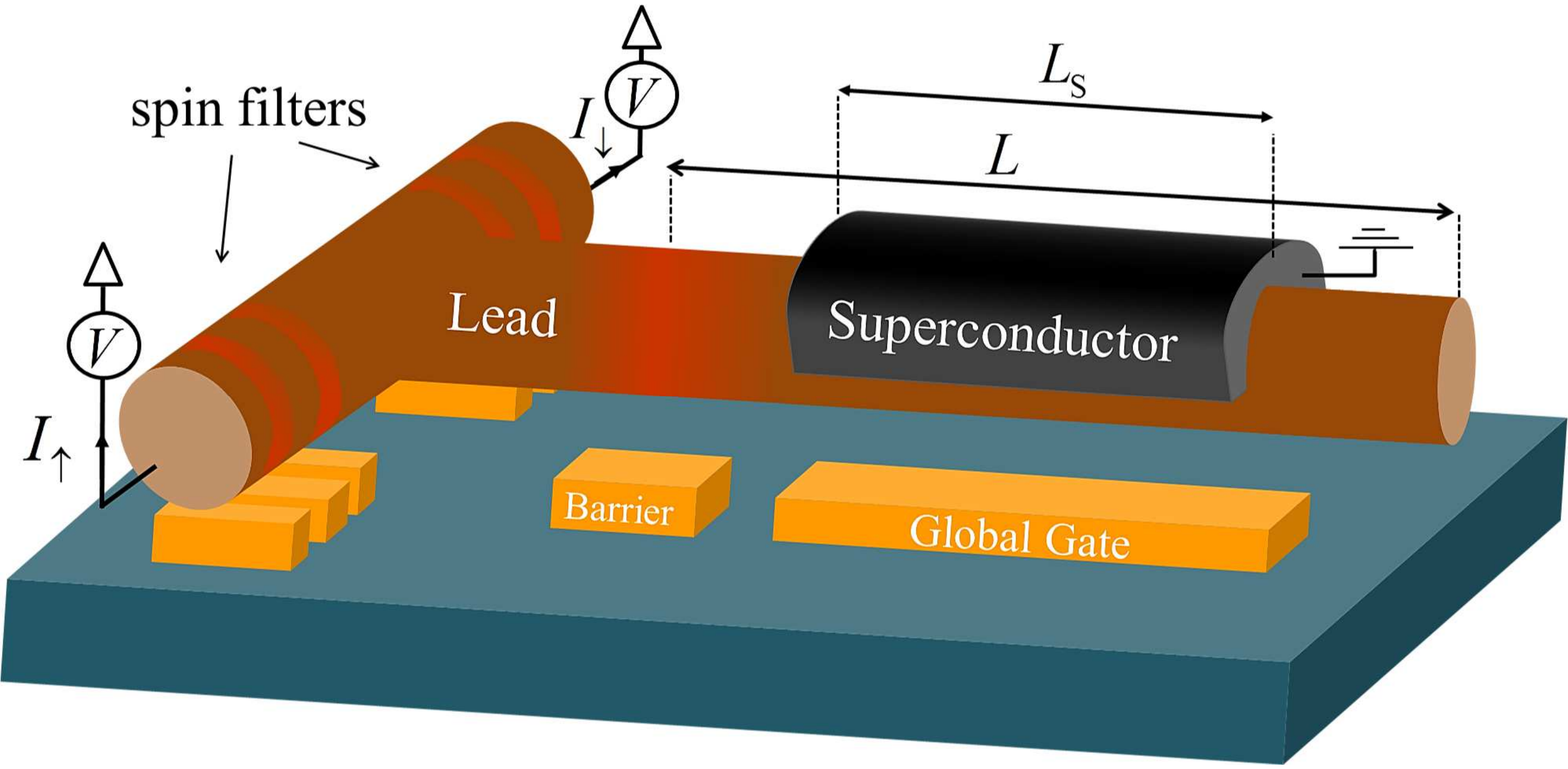}
\caption{A semiconductor nanowire proximity coupled to an {\it s}-wave superconductor. Under certain conditions the system hosts Majorana bound states at its ends. The system is tunnel coupled on the left to a normal lead which is biased at a voltage $V$. The correlations between the spin-resolved currents ($I_{\uparrow}$ and $I_\downarrow$) in the normal lead have features which are unique to the Majorana bound state. To measure these correlations we suggest implementing the system in a $T$-shaped junction and placing a ``spin filter" at each of the arms of the $T$. This may be done by defining quantum dots using gate voltages. In the presence of a magnetic field the resonance level of each dot can be tuned by back gates to have opposite spins.
}\label{fig:setup}
\end{figure}

In this Letter, we discuss the signatures of MBSs in spin-resolved current correlations. Consider a normal metallic lead coupled to a topological superconductor with a MBS at its end. A bias voltage is applied between the lead and the superconductor, driving a current from the lead.
We are interested in the spin-resolved current correlations in the lead, defined as
\begin{equation}
\begin{matrix}
P_{ss'}=\displaystyle\int_{-\infty}^\infty dt\langle\delta \hat{I}_s(0)\delta \hat{I}_{s'}(t)\rangle,&
\delta \hat{I}_s = \hat{I}_s-\langle\hat{I}_s\rangle
\end{matrix},\label{eq:noise_definition}
\end{equation}
where $\hat{I}_s$ is the spin-$s$ current operator ($s = \uparrow,\downarrow$). We concentrate on the cross correlation term~$P_{\uparrow \downarrow}$ and compare between two cases: with a MBS present, and with an ``accidental'' low-energy Andreev bound state (ABS) but without a MBS. Both cases lead to a similar ZBCP.

As we will show, in the presence of a MBS, the cross term~$P_{\uparrow \downarrow}$ carries unique signatures, that are strikingly different from the case of an ABS: In the MBS case, $P_{\uparrow \downarrow}$ is negative in sign, and approaches zero as $P_{\uparrow \downarrow} \propto 1/V$ with increasing bias voltage $V$. In contrast, an ABS generically gives rise to a positive $P_{\uparrow \downarrow}$, that approaches a nonzero constant at high voltages.
Notice that a low-energy ABS can be viewed as a pair of overlapping MBSs. Crucially, however, unlike the case of two spatially separated MBSs, in this case both MBSs are coupled to the lead with comparable strengths.

As a prototypical setup for measuring this effect, we consider a long semiconductor nanowire proximity coupled to a conventional bulk {\it s}-wave superconductor (Fig.~\ref{fig:setup}). Under the right conditions, a MBS is formed at the end of the wire~\cite{lutchyn2010majorana,oreg2010helical}. The wire is tunnel coupled to a normal lead forming a $T$ junction; a bias voltage is applied between the lead and the superconductor. At the two arms of the $T$ junction, there are ``spin filters'' that allow only electrons of a certain spin polarization to pass. Physical ways to implement such spin filters will be discussed below. This setup allows measurement of the correlation functions defined in Eq.~(\ref{eq:noise_definition}), including $P_{\uparrow\downarrow}$ for oppositely polarized spin filters.

\emph{Intuitive analysis.---}The behavior of $P_{\uparrow \downarrow}$ can be understood from qualitative considerations. We assume that the bias voltage is smaller than the gap of the superconductor, such that only Andreev reflection~\cite{andreev1964thermal} in the lead contributes to the conductance. Consider first the case $V \gg \Gamma$, where $\Gamma$ is the width of the ZBCP (originating either from a MBS or an accidental low-energy ABS near the end of the wire). In this limit one can discuss sequential single-particle tunneling events. As Cooper pairs are transported from the superconductor to the lead, they split such that one electron goes to the lead while the other electron changes the occupation of the low-energy resonance~\cite{changing_BS_occu}.

A change in the occupation number of an ordinary ABS is generically accompanied by a change in local physical observables near the edge, e.g., the local spin and charge densities. As a result, the spin density near the edge changes each time an electron is transmitted and the spin of the transmitted electron tends to be antialigned with the spin of the preceding transmitted electron. (If the $z$ component of the spin is conserved, this correlation is perfect.) Such a correlation corresponds to $P_{\uparrow \downarrow}>0$.

On the other hand, a change in the occupation number of a MBS cannot be detected in \emph{any} local observable near a single edge. In particular, the local spin densities of the two degenerate ground states (associated with the occupation number of the fermion formed by the MBSs at the two ends) are identical~\cite{equal_spin_density}. It follows that the spins of consecutive electrons are uncorrelated; hence~$P_{\uparrow \downarrow} \rightarrow 0$.

Next, consider the case $V, T \ll \Gamma$ (where $T$ is the temperature). In the MBS case, the total shot-noise power $P=\sum_{s,s'=\uparrow,\downarrow}P_{ss'}$ goes to zero as a result of the total transmission approaching unity~\cite{Law2009majorana,Bolech2007Observing,Fidkowski2012universal}. Since $P_{\uparrow\uparrow}$ and $P_{\downarrow\downarrow}$ are positive definite quantities, we must have $P_{\uparrow\downarrow}=P_{\downarrow\uparrow}\le0$.

\emph{Simple model.---}With this qualitative picture in mind, we calculate $P_{ss'}$ for a general low-energy model $H = H_L + H_T$ of a normal lead coupled to a MBS, where
\begin{equation}
\begin{array}{cc}
H_L=\displaystyle\sum_{k,s}\displaystyle \epsilon_k\psi_{k s}^\dag\psi^{\phantom \dag}_{k s},&H_T = i\gamma\cdot\displaystyle\sum_{k,s}\displaystyle (t_s^{\phantom *}\psi^{\phantom \dag}_{k s}+{\rm H.c.}).
\end{array}\label{eq:H_MBS}
\end{equation}
Here $\psi_{ks}$ describes the lead modes with spin $s$, $\epsilon_k$ are the energy levels in the lead, and $t_s$ is the coupling constant of these modes to the Majorana state described by $\gamma$~\cite{multi_channel_2}. The form of $H$ is quite general and stems solely from the Hermitian nature of $\gamma$.

At energies below the superconducting gap only reflection processes are possible, and the scattering matrix is given by~\cite{Fisher1981relation,Iida1990statistical}
\begin{equation}
\begin{pmatrix}r^{ee}&r^{eh}\\r^{he}&r^{hh}\end{pmatrix}=1-2\pi iW^\dag\left(E+i\pi WW^\dag\right)^{-1}W ,\label{eq:Weidenmuller}
\end{equation}
with $W=\sqrt{\nu_0}(t_\uparrow,t_\downarrow,t_\uparrow^\ast,t_\downarrow^\ast)$, and where $\nu_0$ is the density of states in the lead. This yields
\begin{equation}
\begin{array}{ccc}
\mathlarger{r_{ss'}^{ee}=\delta_{ss'}+\frac{2\pi \nu_0 t_s^\ast t_{s'}}{iE-\Gamma}}&,&
\mathlarger{r_{ss'}^{he}=\frac{2\pi \nu_0 t_s t_{s'}}{iE-\Gamma}}
\end{array}\label{eq:S_mat_MBS}
\end{equation}
where $r^{hh}\mathsmaller{(E)}=[r^{ee}\mathsmaller{(-E)}]^\ast$, $r^{eh}\mathsmaller{(E)}=[r^{he}\mathsmaller{(-E)}]^\ast$ as dictated by particle-hole symmetry, and $\Gamma=2\pi \nu_0 (|t_\uparrow|^2+|t_\downarrow|^2)$.

The spin-resolved currents and their correlation functions are given by~\cite{Anantram1996current}
\begin{equation}
\begin{split}
\langle \hat{I}_s\rangle &= \frac{e}{h}\displaystyle\sum_{\renewcommand{\arraystretch}{0.3}\begin{array}{c}\mathsmaller{
s'\in\uparrow,\downarrow}\\\mathsmaller{\alpha,\beta\in e,h}\end{array}}\sgn(\alpha) \displaystyle\int_0^\infty dE A^{\beta\beta}_{s's'}(s,\alpha;E)f_\beta(E),\\
P_{ss'}& = \frac{e^2}{h}\displaystyle\sum_{ \renewcommand{\arraystretch}{0.3}\begin{array}{c}\mathsmaller{\sigma,\sigma'\in\uparrow,\downarrow}\\
\mathsmaller{\alpha,\beta,\gamma,\delta\in e,h}\end{array}
}\sgn(\alpha)\sgn(\beta)\displaystyle\int_0^\infty dE\\  \times& A^{\gamma\delta}_{\sigma\sigma'}(s,\alpha;E)A^{\delta\gamma}_{\sigma'\sigma}(s',\beta;E) f_\gamma(E)[1-f_\delta(E)],\\
A^{\gamma\delta}_{\sigma\sigma'}&(s,\alpha;E)= \delta_{s\sigma}\delta_{s\sigma'}\delta_{\alpha\gamma}\delta_{\alpha\delta}- [r^{\alpha\gamma}_{s\sigma}]^\ast r^{\alpha\delta}_{s\sigma'},
\end{split}\label{eq:noise_formula}
\end{equation}
with $f_e(E)=1-f_h(-E)$ being the distribution of incoming electrons in the lead. Here $\sgn(\alpha)=+1$ for $\alpha=e$ and $\sgn(\alpha)=-1$ for $\alpha=h$. Inserting the reflection matrices of Eq.~\eqref{eq:S_mat_MBS}, one obtains at zero temperature
\begin{equation}
P_{\uparrow\downarrow}=-\frac{2e^2}{h}\Gamma_\uparrow\Gamma_\downarrow \frac{eV}{(eV)^2+\Gamma^2},\label{eq:MBS_noise}
\end{equation}
where $\Gamma_s=2\pi \nu_0 |t_s|^2$. As anticipated, $P_{\uparrow\downarrow}$ is negative and goes to zero at high voltages as $1/V$ (assuming $eV$ remains smaller than the superconducting gap). We note that summing $P_{\uparrow\downarrow}$ with the rest of the spin-resolved terms gives the total shot noise power~\cite{Golub2011shot,golub2014interferometric,Nilsson2008splitting,Bolech2007Observing}.

The result of Eq.~\eqref{eq:MBS_noise} does not depend on details such as the particular system hosting the MBS, the nature of the coupling to the lead, or the particular spin polarization axis. One can change the spin axis by transforming the coupling constants according to
\begin{equation}
\begin{pmatrix}t'_\uparrow,&t'_\downarrow\end{pmatrix}= \begin{pmatrix}t_\uparrow,&t_\downarrow\end{pmatrix}\cdot\exp(-i\theta\hat{\mathbf{n}}\cdot\boldsymbol{\sigma}/2),
\end{equation}
where $\boldsymbol{\sigma}$ is a vector of Pauli matrices, $\hat{\bf{n}}$ is a unit vector and $\theta$ is a rotation angle. We note in passing that by varying both $\hat{\bf{n}}$ and $\theta$ one can always find a spin axis such that $t'_\downarrow=0$~\cite{Leijnse2011quantum}, resulting in a spin-polarized current~\cite{He2014selective}.

Next, we consider an accidental low-energy ABS. For simplicity we shall temporarily assume that spin in the $z$ direction is conserved~\cite{Footnote}. Under these assumptions the most general tunneling Hamiltonian is given by~\cite{SM}
\begin{equation}
\tilde{H}_T = a^\dag\displaystyle\sum_{k}\left(\tilde{t}_\uparrow\psi_{k \uparrow}+ \tilde{t}_\downarrow\psi^\dag_{k\downarrow}\right) + {\rm H.c.} ,
\end{equation}
where $a$ is the annihilation operator for the ABS. (Notice that if one writes $a$ in terms of two Majorana operators, then both of them are coupled to the lead with equal strength.) One can now use Eq.~\eqref{eq:Weidenmuller} with
\begin{equation}
W=\sqrt{\nu_0}\begin{pmatrix}\tilde{t}_\uparrow &0&0&\tilde{t}^\ast_\downarrow\\ 0& \tilde{t}_\downarrow& \tilde{t}^\ast_\uparrow&0 \end{pmatrix} ,
\end{equation}
to obtain the reflection matrices
\begin{equation}
\begin{array}{ccc}
r^{ee}=\frac{iE}{iE-\tilde\Gamma/2}+\frac{(\tilde\Gamma_{\uparrow}-\tilde\Gamma_{\downarrow})/2}{iE-\tilde\Gamma/2}\sigma^z&,&
r^{he}= \frac{2\pi \nu_0 \tilde{t}_{\uparrow}\tilde{t}_{\downarrow}}{iE-\tilde\Gamma/2}\sigma^x
\end{array}.\label{eq:S_mat_ABS}
\end{equation}

These reflection matrices are written in the basis of the spin in the $z$ direction. To obtain them for a general spin direction, we perform a transformation on $r^{ee}$, $r^{he}$ which rotates the spin axis by an angle $\theta$ away from the $z$ axis~\cite{SM}.
Upon doing so, and then using Eq.~\eqref{eq:noise_formula} one has
\begin{equation}
\begin{split}
P_{\uparrow\downarrow}=\mathsmaller{\frac{2e^2}{h}\frac{\tilde\Gamma_\uparrow\tilde\Gamma_\downarrow}{\tilde\Gamma}}
\Big\{&[\mathsmaller{\frac{(\tilde\Gamma_\uparrow-\tilde\Gamma_\downarrow)^2}{\tilde\Gamma^2}+\cos^2\theta}] \cdot\arctan\mathsmaller{\frac{2eV}{\tilde\Gamma}}\\
\mathsmaller{+}& [\mathsmaller{\frac{(\tilde\Gamma_\uparrow-\tilde\Gamma_\downarrow)^2}{\tilde\Gamma^2}-\cos^2\theta}] \mathsmaller{\cdot\frac{2eV/\tilde\Gamma}{1+(2eV/\tilde\Gamma)^2}} \Big\} .
\end{split}\label{eq:ABS_noise}
\end{equation}
This should be compared to Eq.~\eqref{eq:MBS_noise}. Unlike the MBS scenario, $P_{\uparrow\downarrow}$ is now positive for all $V$ and monotonically approaches a finite value at $eV\gg\tilde\Gamma$.

\emph{Microscopic model.---}Next we verify our conclusions using a numerical simulation of an experimentally realizable microscopic model~\cite{mourik2012signatures,Deng2012a,Das2012zero,churchill2013superconductor}. We consider a nanowire having Rashba spin-orbit coupling, proximity coupled to an {\it s}-wave superconductor, with an applied Zeeman field. The wire is tunnel coupled to a normal lead from the left, as depicted in Fig.~\ref{fig:setup}. The Hamiltonian for the system (not including the spin filters) is $H=H_L+H_{\rm nw}+H_T$, with $H_L$ being the isolated lead Hamiltonian in Eq.~\eqref{eq:H_MBS}, $H_{\rm nw}$ is the Hamiltonian for the nanowire given by
\begin{equation}
\begin{split}
&H_{\rm nw}=\displaystyle\int_{-L/2}^{L/2} dx\Phi^\dag(x)\mathcal{H}\Phi(x),\\
\mathcal{H}= (\frac{-\partial^2_x}{2m_{\rm e}}&-\mu)\tau^z+i\alpha_R\tau^z\sigma^z\partial_x+\mathbf{B}\mathsmaller{\cdot}\boldsymbol{\sigma}+\Delta(x)\tau^x,
\end{split}\label{eq:NW_Hamiltonian}
\end{equation}
where $\Phi^\dag(x)=(\phi^{\mathsmaller{\dag}}_{\mathsmaller{\uparrow}}(x),\phi^{\mathsmaller{\dag}}_{\mathsmaller{\downarrow}}(x), \phi_{\mathsmaller{\downarrow}}(x),\mathsmaller{-}\phi_{\mathsmaller{\uparrow}}(x))$ are the electron creation and annihilation operators in the wire, and $H_T$ describes the coupling of the nanowire to the lead
\begin{equation}
H_T =-\displaystyle\sum_{k,p,s}t_{kp}\phi^\dag_{ps}\psi_{ks} +{\rm H.c.} .
\end{equation}
Here $\phi_{ps}$ denotes an eigenmode of the decoupled wire, $t_{kp}$ are the hopping matrix elements between the lead and the wire, $m_{\rm e}$ is the effective electron mass, $\mu$ is the chemical potential, $\alpha_R$ describes the spin-orbit coupling, $\mathbf{B}$ is the Zeeman field, and $\Delta(x)=\Delta_0\theta(L_{\rm S}/2-|x|)$ is the induced pair potential in the wire, with $L_{\rm S}$ being the length of the section of the wire which is covered by the superconductor (cf. Fig.~\ref{fig:setup}).

As we shall now show, this system can exhibit either a zero-energy ABS or a zero-energy MBS at the end of the wire, depending on the value of $B$. The differential conductance spectra in the two cases are similar. The spin-resolved current correlations, however, are qualitatively different. By discretizing $H$ on a lattice we numerically obtain the scattering matrix~\cite{SM}, from which the spin-resolved currents and their correlations are obtained with the help of Eq.~\eqref{eq:noise_formula}.
\begin{figure}
\begin{tabular}{@{\hskip -0.15cm}l @{\hskip -0.075cm}r}
\,\,\includegraphics[clip=true, trim =3.5cm 9.28cm 4.2cm 10.1cm,width=0.245\textwidth]{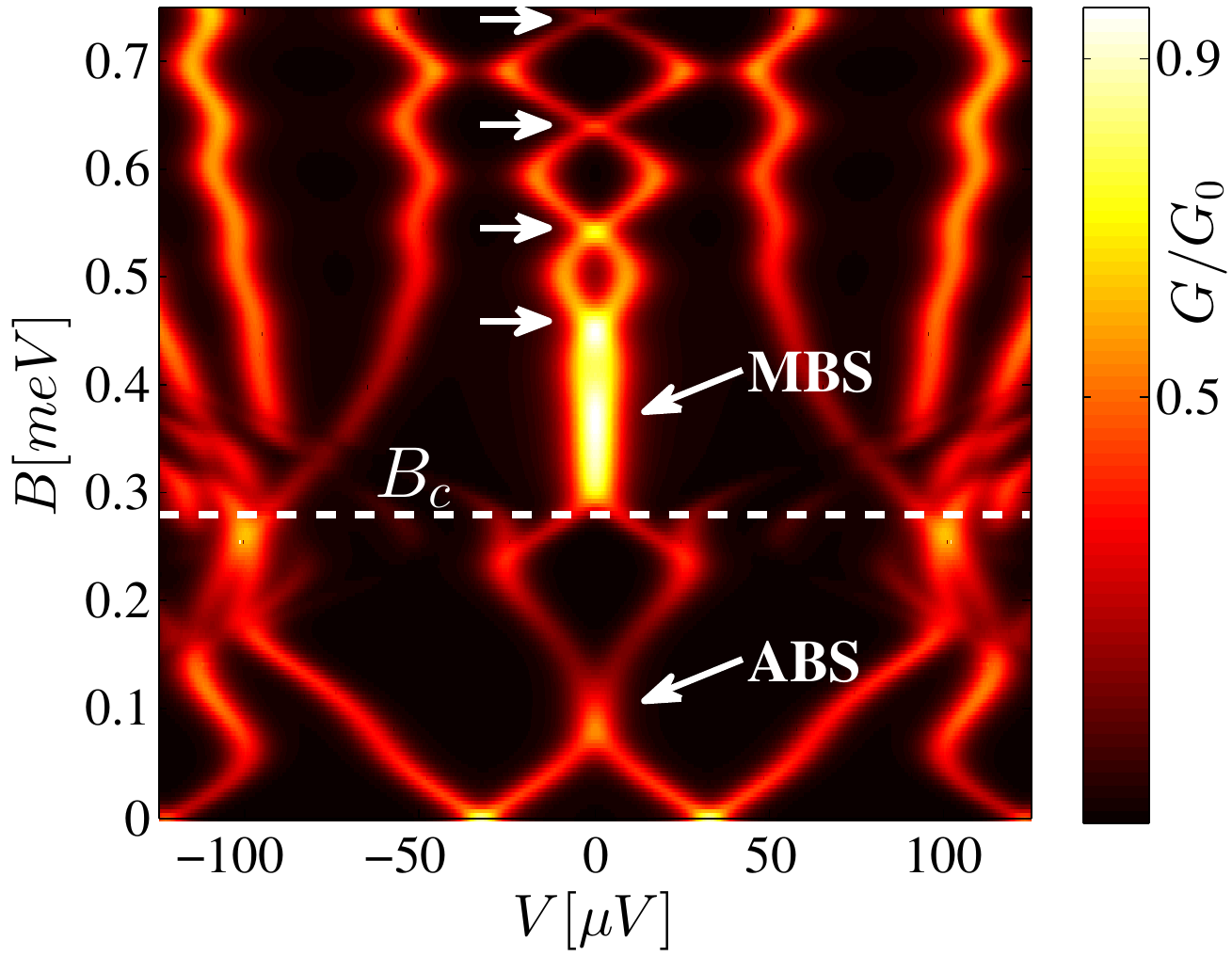} \llap{\parbox[c]{8.4cm}{\vspace{-1mm}\footnotesize{(a)}}} &
\includegraphics[clip=true, trim =1.1cm -0.3cm 1cm 0.8cm,width=0.23\textwidth]{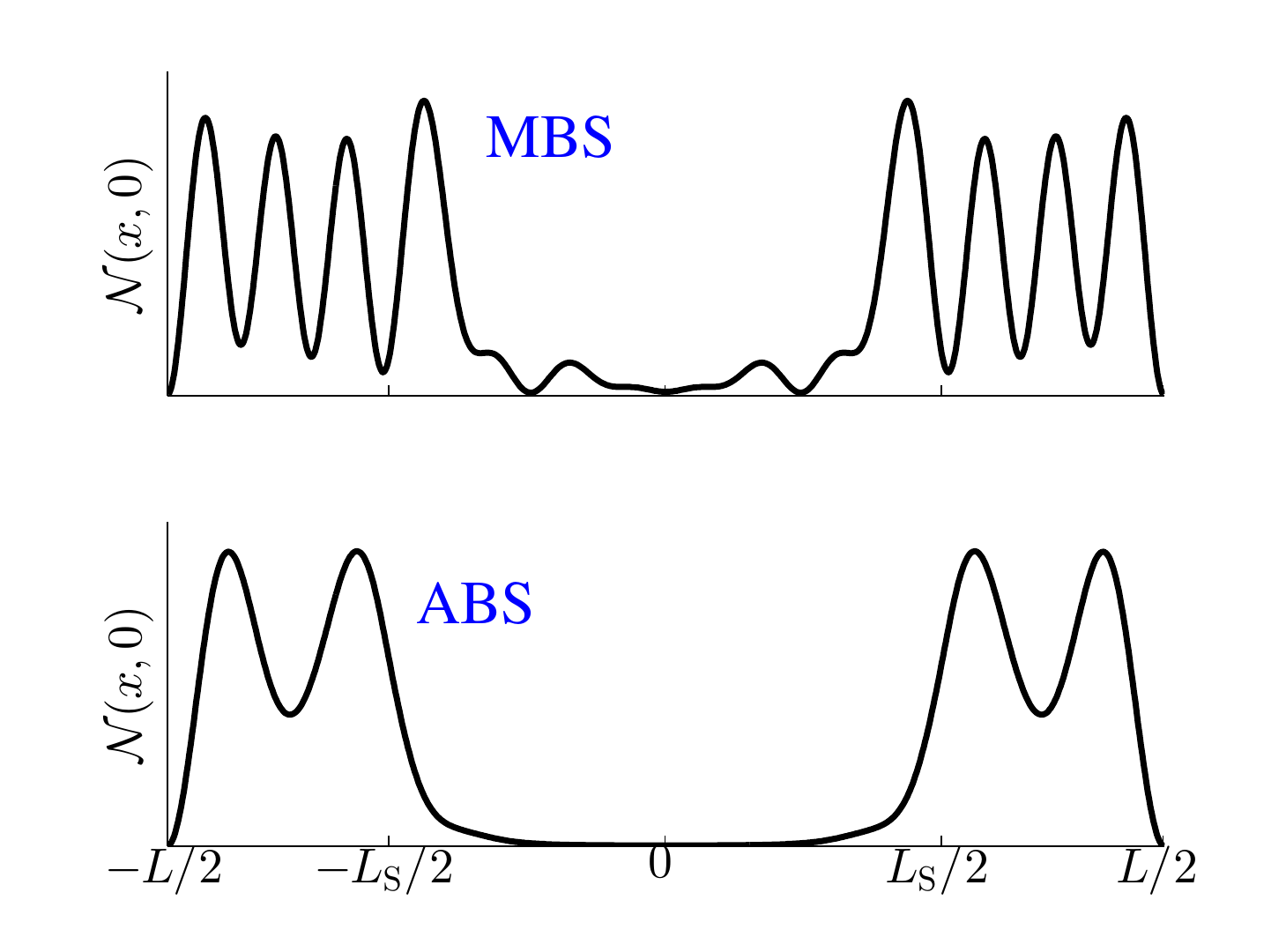} \llap{\parbox[c]{8cm}{\vspace{-1mm}\footnotesize{(b)}}} \\[-0.25ex]
\,\,\,\includegraphics[clip=true, trim =1.57cm 0cm 1.1cm 0.6cm,width=0.232\textwidth]{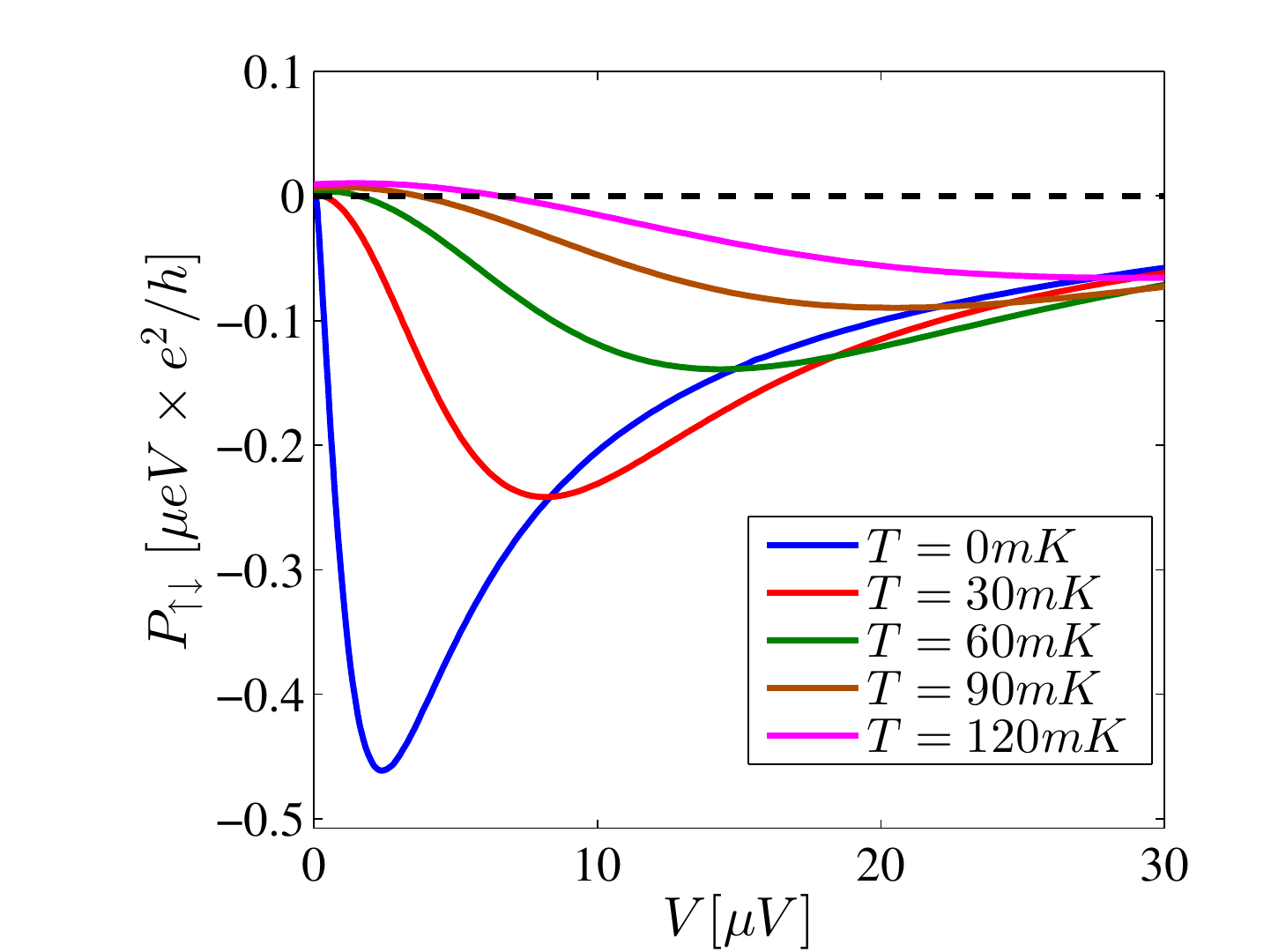} \llap{\parbox[c]{8cm}{\vspace{-1.5mm}\footnotesize{(c)}}} &
\,\,\includegraphics[clip=true, trim =1.6cm 0cm 1.1cm 0.8cm,width=0.229\textwidth]{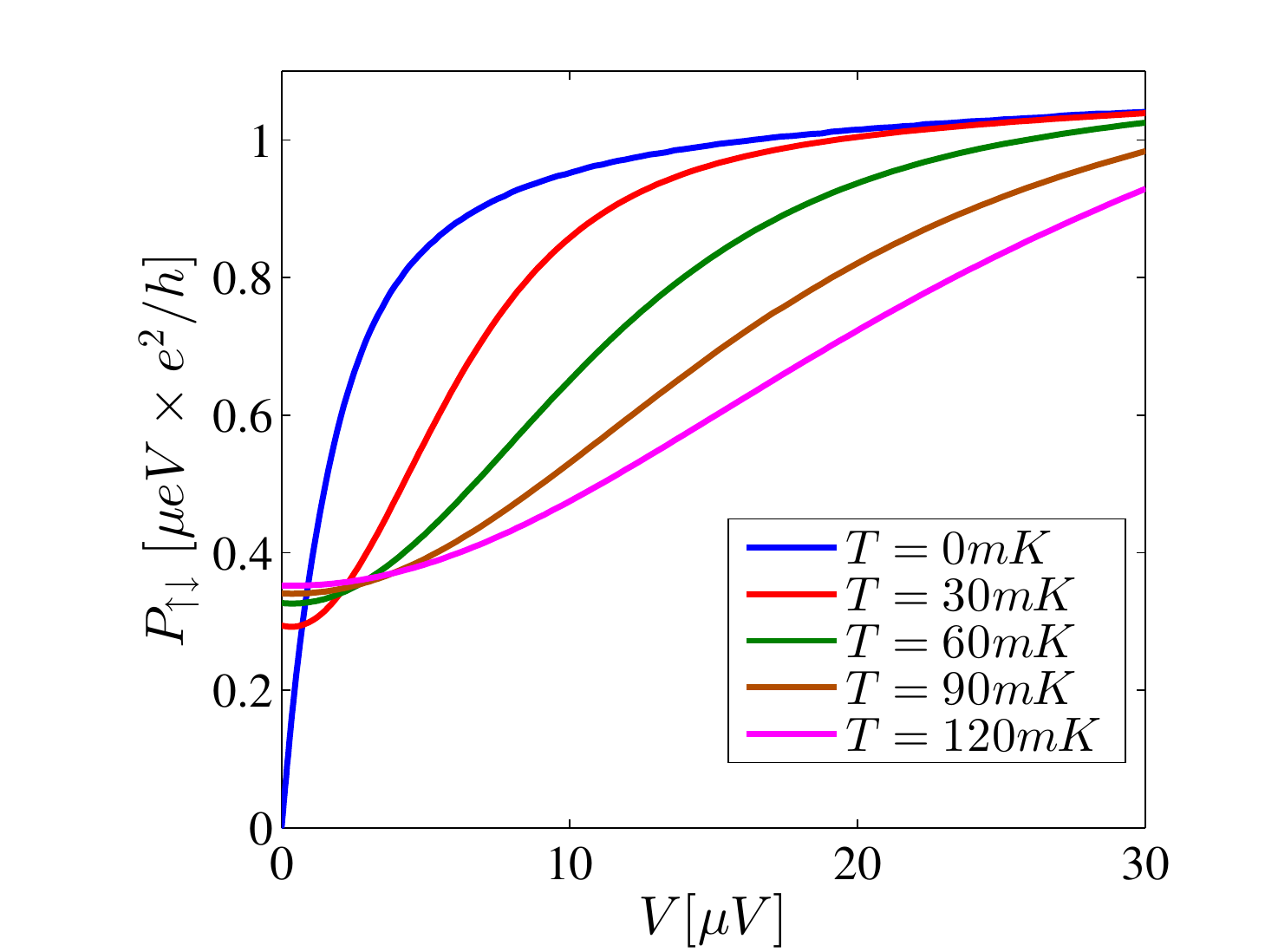} \llap{\parbox[c]{8cm}{\vspace{-1.5mm}\footnotesize{(d)}}}
\end{tabular}
\caption{Numerical simulation of the system described in Eq.~\eqref{eq:NW_Hamiltonian} and depicted in Fig.~\ref{fig:setup}. The parameters of the system are taken to be in accordance with a recent experiment~\cite{mourik2012signatures}, namely $E_{\rm so}=m_{\rm e}\alpha_R^2/2=50\mu eV$, $\Delta_0=250\mu eV$, and $l_{\rm so}=1/(m_{\rm e}\alpha_R)=200nm$. We take the length of the wire to be $L=2.5\mu m$ with $L_{\rm S}=1.4\mu m$. Similar results are obtained for parameters taken from a different experiment~\cite{Das2012zero,SM}. The magnetic field $\mathbf{B}$ is applied at an angle of $60^\circ$ from the $z$ axis in the $xz$ plane. (a) Differential conductance as a function of bias $V$ and Zeeman energy $B$ for $\mu=125\mu eV$ and at $T=30mK$, in units of $G_0=e^2/h$. A zero-bias conductance peak appears both as a result of a Majorana bound state (MBS) at $B>B_c$, and as a result of a trivial Andreev bound state (ABS) at $B<B_c$. (b) Local density of state at zero energy for $B=350\mu eV$ and for $B=90\mu eV$, where the system hosts a localized MBS and an ABS, respectively. In both cases the density of states is significant only near the ends of the wire. (c),(d) Spin-resolved currents correlation $P_{\uparrow\downarrow}$ vs $V$ at different temperatures for (c) the MBS and (d) the ABS. For the Majorana case, $P_{\uparrow\downarrow}$ is negative and goes to zero at large $V$. This is in striking contrast to the case of an ABS, where $P_{\uparrow\downarrow}$ is positive and approaches a finite constant value at large $V$.}\label{fig:dI_dV_LDOS_P}
\end{figure}

In Fig.~\hyperref[fig:dI_dV_LDOS_P]{\ref{fig:dI_dV_LDOS_P}a} the differential conductance $d\langle \hat{I}\rangle/dV$ is presented as a function of bias voltage $V$ and Zeeman energy $B$, for a value of $\mu=125\mu eV$ and at a temperature of $T=30mK$.
The magnetic field $\mathbf{B}$ is applied at an angle of $60^\circ$ from the $z$ axis in the $xz$ plane.
The dashed white line signifies the critical Zeeman energy $B_c=\sqrt{\mu^2+\Delta_0^2}$ above which the system is in the topological phase in the thermodynamic limit~\cite{lutchyn2010majorana,oreg2010helical}. Beyond this a zero-energy MBS appears, and one observes a ZBCP. At even higher magnetic fields the conductance begins to oscillate due to the overlap between the MBSs at the two ends of the wire~\cite{Lim2012magnetic,Das-Sarma2012splitting,Das2012zero,Rainis2013Towards}.

Importantly, a ZBCP is also present at a magnetic field which is \emph{below} the critical line, at about $B\sim0.1meV$, even though the system is in the topologically trivial phase.  This ZBCP is due to a trivial ABS which is localized at the left end of the wire. In Fig.~\hyperref[fig:dI_dV_LDOS_P]{\ref{fig:dI_dV_LDOS_P}b} the local density of states (LDOS) at zero energy $\mathcal{N}(x,0)$~\cite{SM} is presented for two different Zeeman energies $B=350\mu eV$ and $B=90\mu eV$, corresponding to the MBS and ABS, respectively. We note that in both cases the LDOS is peaked at the two ends of the wire~\cite{Klinovaja2012Composite}, making it difficult to distinguish between the ABS and the MBS via a scanning tunneling microscopy measurement.

The spin-resolved current correlation $P_{\uparrow\downarrow}$, on the other hand, is qualitatively different for the two cases. Figure \hyperref[fig:dI_dV_LDOS_P]{\ref{fig:dI_dV_LDOS_P}c} and Fig.~\hyperref[fig:dI_dV_LDOS_P]{\ref{fig:dI_dV_LDOS_P}d} show $P_{\uparrow\downarrow}$ as a function of bias for the MBS ($B=350\mu eV$) and for the ABS ($B=90\mu eV$), respectively. As anticipated, in the case of a MBS the correlations are negative and approach zero at high voltages. In the case of an ABS, the correlations are positive and approach a finite value at large $V$. This is in agreement with the analytical low-energy treatment which resulted in Eq.~\eqref{eq:MBS_noise} and Eq.~\eqref{eq:ABS_noise}.

Interestingly, the main features distinguishing a MBS from an ABS survive even at finite temperatures, as apparent in Figs.~\hyperref[fig:dI_dV_LDOS_P]{\ref{fig:dI_dV_LDOS_P}c},~\hyperref[fig:dI_dV_LDOS_P]{\ref{fig:dI_dV_LDOS_P}d}. At a finite temperature, $P_{\uparrow \downarrow}\ne 0$ at zero voltage. $P_{\uparrow\downarrow}$ recovers its low-$T$ behavior at voltage $V \gtrsim T$. In particular, one can witness these distinctive features even for $T>\Gamma$. We note that Figs.~\hyperref[fig:dI_dV_LDOS_P]{\ref{fig:dI_dV_LDOS_P}c},~\hyperref[fig:dI_dV_LDOS_P]{\ref{fig:dI_dV_LDOS_P}d} present results for voltages that are smaller than the excitation gap in the system (roughly $50\mu eV$). At higher voltages the features of $P_{\uparrow\downarrow}$ are no longer universal as $P_{\uparrow\downarrow}$ picks up contributions from higher-energy resonances~\cite{SM}.

\begin{figure}
\begin{tabular}{@{\hskip -0.0cm}l @{\hskip -0.0cm}r}
\includegraphics[clip=true, trim =1.55cm 0cm 1.1cm 0.8cm,width=0.236\textwidth]{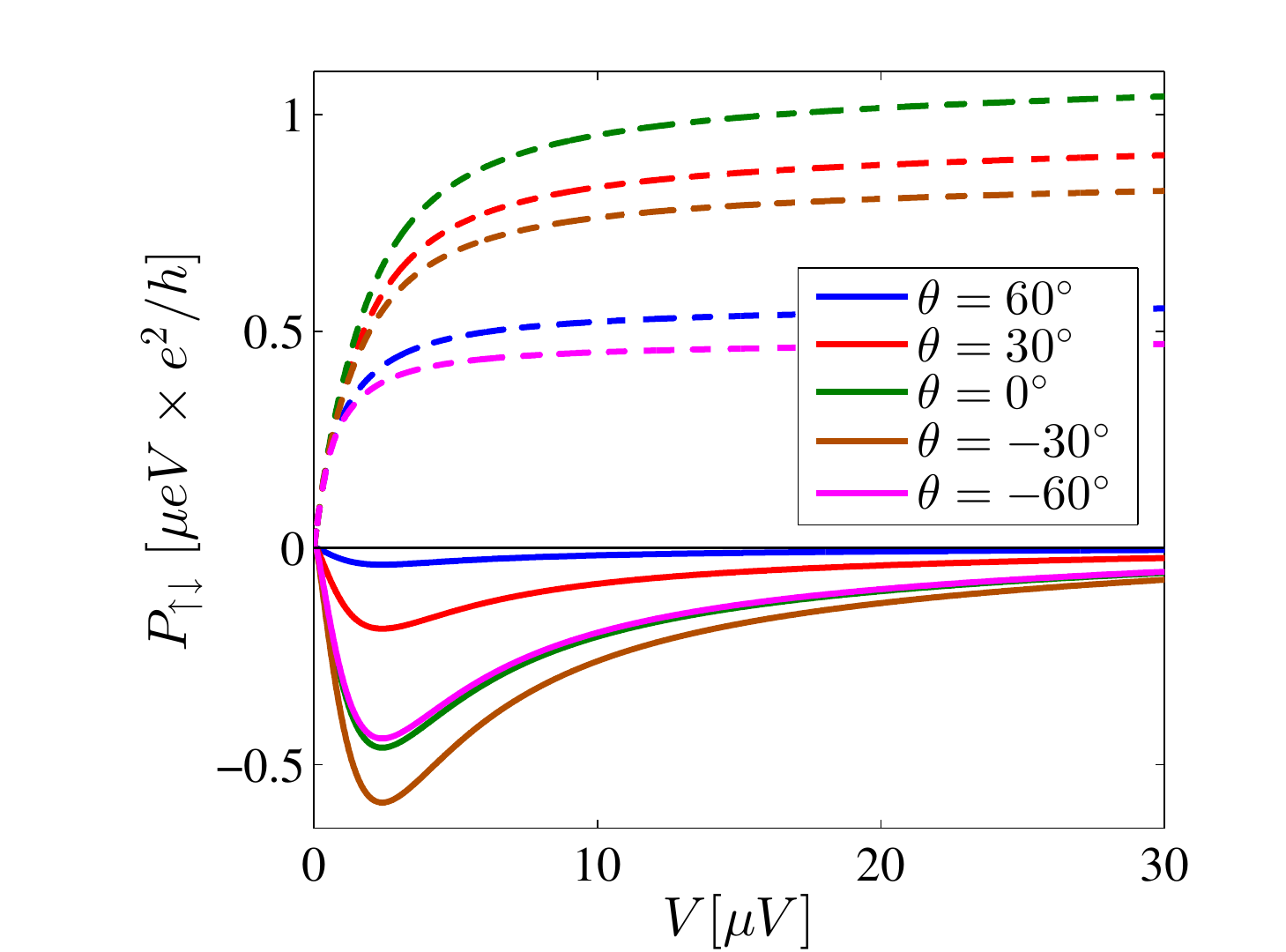} \llap{\parbox[c]{8cm}{\vspace{-1.5mm}\footnotesize{(a)}}} &
\includegraphics[clip=true, trim =1.59cm 0cm 1.1cm 0.6cm,width=0.236\textwidth]{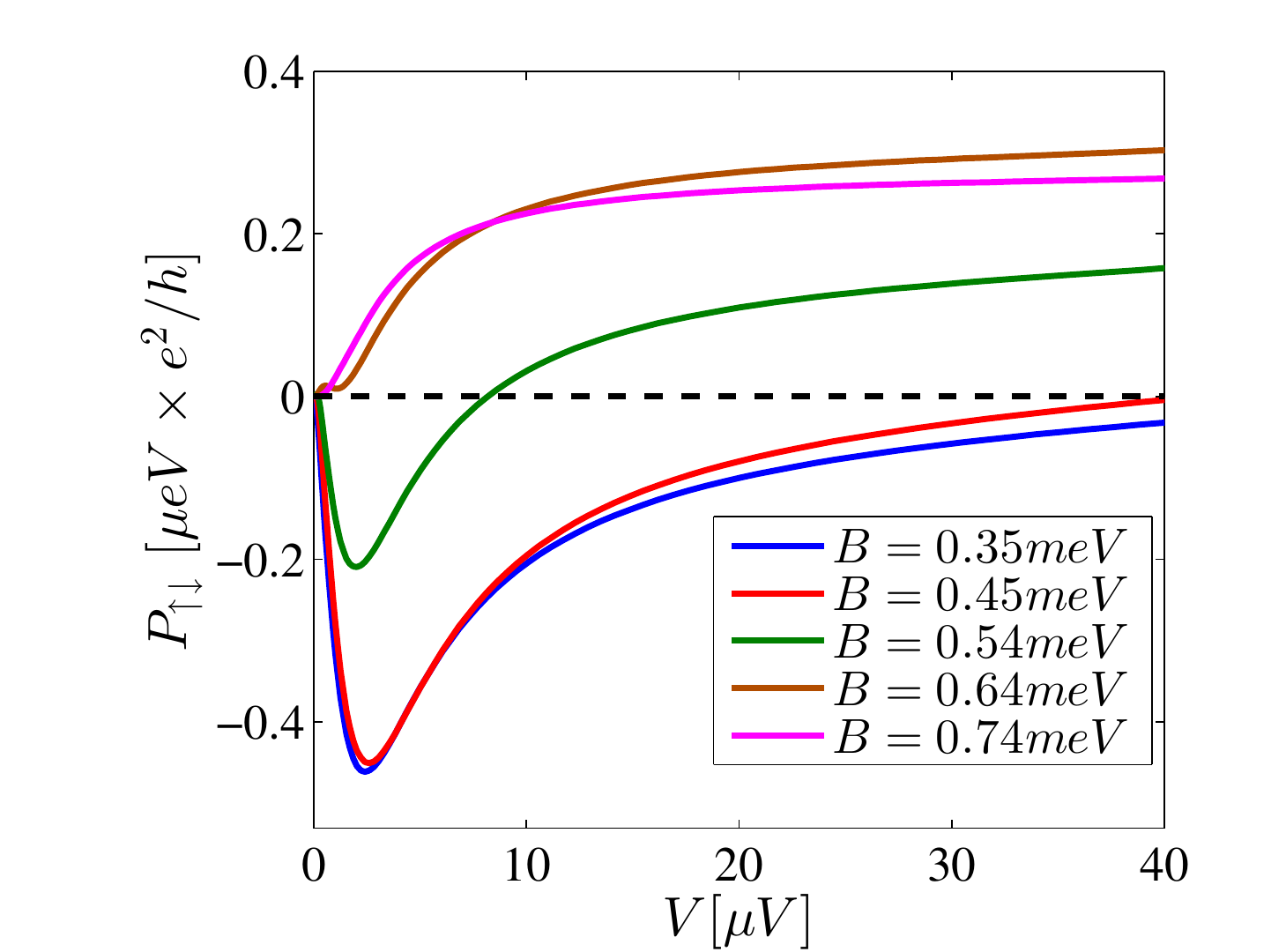} \llap{\parbox[c]{8cm}{\vspace{-1.5mm}\footnotesize{(b)}}}
\end{tabular}
\caption{Spin-resolved current correlations $P_{\uparrow\downarrow}$ as a function of bias voltage $V$, at $T=0$. (a) The spin-resolved currents are defined with respect to an axis which is rotated by an angle $\theta$ from the $z$ axis in the $xz$ plane. The direction of $\mathbf{B}$ remains fixed at an angle of $60^\circ$ from the z axis. The characteristic features  seem to be angle independent for both the Majorana bound state (MBS) $B=350\mu eV$ (solid lines), and the trivial Andreev bound state (ABS), $B=90\mu eV$ (dashed lines). (b) Crossover between a MBS and an ABS. As $B$ is increased the spatial overlap of the pair of Majorana end states increases until they are indistinguishable from an ordinary ABS (cf. marked points in Fig.~\hyperref[fig:dI_dV_LDOS_P]{\ref{fig:dI_dV_LDOS_P}a}).}\label{fig:P_vs_theta_and_CO}
\end{figure}

The spin-resolved currents whose correlation is presented in Figs.~\hyperref[fig:dI_dV_LDOS_P]{\ref{fig:dI_dV_LDOS_P}c},~\hyperref[fig:dI_dV_LDOS_P]{\ref{fig:dI_dV_LDOS_P}d} are all defined with respect to the $z$ spin axis. In Fig.~\hyperref[fig:P_vs_theta_and_CO]{\ref{fig:P_vs_theta_and_CO}a} we present $P_{\uparrow\downarrow}$ for spin-resolved currents defined with a spin axis rotated by an angle $\theta$ from the $z$ axis in the $xz$ plane~\cite{SM}. The results for the MBS (solid lines) and for the ABS (dashed lines) are obtained at zero temperature and for the same parameters as those of Fig.~\hyperref[fig:dI_dV_LDOS_P]{\ref{fig:dI_dV_LDOS_P}c} and Fig.~\hyperref[fig:dI_dV_LDOS_P]{\ref{fig:dI_dV_LDOS_P}d}, respectively. It is apparent that the same distinctive features persist upon rotating the spin axis. We point out the suppression of $P_{\uparrow\downarrow}$ in the MBS case for $\theta=60^\circ$, which is the direction of $\mathbf{B}$. This is caused due to polarization of the Majorana wave function~\cite{Sticlet2012spin,guigou2014signature} in the $\mathbf{B}$ direction, giving rise to a nearly perfect polarization of the spin-resolved current through the MBS.

It is interesting to examine the crossover between the MBS case and the ABS case. This can be done by increasing $B$ to the point where there is a large overlap between the MBSs at the two ends of the wire. At this point, the two Majorana states are equivalent to a single ordinary ABS. In particular, they are both coupled to the lead with comparable strengths. In Fig.~\hyperref[fig:P_vs_theta_and_CO]{\ref{fig:P_vs_theta_and_CO}b} we present $P_{\uparrow\downarrow}$ vs $V$ for various Zeeman energies $B$, corresponding to two MBSs with increasing spatial overlap. As the overlap increases, $P_{\uparrow\downarrow}$ turns from being negative to being positive for all $V$. We note that for all these values of $B$ a ZBCP is present in the differential conductance spectra [cf. Fig.~\hyperref[fig:dI_dV_LDOS_P]{\ref{fig:dI_dV_LDOS_P}a}].

\emph{Discussion.---}We have shown that a MBS has unique signatures in spin-resolved current correlations, distinguishing it from a topologically trivial ABS. These signatures are rooted in the nonlocal nature of the MBS. We expect other low-energy resonances, such as a Kondo resonance~\cite{Cuevas2001kondo,Lee2012zero-bias,Cheng2014interplay}, or end modes due to smooth confinement in a nontopological state~\cite{kells2012near}, to behave qualitatively like an ABS.

Finally, we discuss the proposed realization of the spin filters described in Fig.~\ref{fig:setup}. Gates located underneath each of the two normal legs of the junction define two quantum dots. By varying the gate potential under the dot, one can tune a level of a certain spin to be at resonance, thereby filtering the spin-resolved current through that leg~\cite{FN_2}. If the two dots are tuned to opposite spin resonances, $P_{\uparrow\downarrow}$ can be obtained by measuring correlations between the currents through the two normal legs. Alternatively, spin filters can be constructed by coupling the normal legs to oppositely polarized ferromagnets~\cite{He2014selective} or to a quantum spin Hall insulator~\cite{Posske2014direct,Kane2005quantum,Bernevig2006quantum,konig2007quantum}.

\emph{Acknowledgements-} We would like to acknowledge C. W. J. Beenakker, A. Yacoby, M. Heiblum, B. I. Halperin, L. Fu, J. D. Sau, P. W. Brouwer, F. Pientka, I. C. Fulga, Y. Schattner, A. Keselman, E. Sagi, D. Mark and K. Kaasbjerg. This study was supported by the Israel Science Foundation (ISF), a Career Integration Grant (CIG), the German-Israeli Foundation (GIF), the Minerva Foundation, the Helmholtz Virtual Institute - ``New states of matter and their excitations", and an ERC grant (FP7/2007-2013) 340210.

\bibliography{SRCC_References}

\begin{thebibliography}{59}%
\makeatletter
\providecommand \@ifxundefined [1]{%
 \@ifx{#1\undefined}
}%
\providecommand \@ifnum [1]{%
 \ifnum #1\expandafter \@firstoftwo
 \else \expandafter \@secondoftwo
 \fi
}%
\providecommand \@ifx [1]{%
 \ifx #1\expandafter \@firstoftwo
 \else \expandafter \@secondoftwo
 \fi
}%
\providecommand \natexlab [1]{#1}%
\providecommand \enquote  [1]{``#1''}%
\providecommand \bibnamefont  [1]{#1}%
\providecommand \bibfnamefont [1]{#1}%
\providecommand \citenamefont [1]{#1}%
\providecommand \href@noop [0]{\@secondoftwo}%
\providecommand \href [0]{\begingroup \@sanitize@url \@href}%
\providecommand \@href[1]{\@@startlink{#1}\@@href}%
\providecommand \@@href[1]{\endgroup#1\@@endlink}%
\providecommand \@sanitize@url [0]{\catcode `\\12\catcode `\$12\catcode
  `\&12\catcode `\#12\catcode `\^12\catcode `\_12\catcode `\%12\relax}%
\providecommand \@@startlink[1]{}%
\providecommand \@@endlink[0]{}%
\providecommand \url  [0]{\begingroup\@sanitize@url \@url }%
\providecommand \@url [1]{\endgroup\@href {#1}{\urlprefix }}%
\providecommand \urlprefix  [0]{URL }%
\providecommand \Eprint [0]{\href }%
\providecommand \doibase [0]{http://dx.doi.org/}%
\providecommand \selectlanguage [0]{\@gobble}%
\providecommand \bibinfo  [0]{\@secondoftwo}%
\providecommand \bibfield  [0]{\@secondoftwo}%
\providecommand \translation [1]{[#1]}%
\providecommand \BibitemOpen [0]{}%
\providecommand \bibitemStop [0]{}%
\providecommand \bibitemNoStop [0]{.\EOS\space}%
\providecommand \EOS [0]{\spacefactor3000\relax}%
\providecommand \BibitemShut  [1]{\csname bibitem#1\endcsname}%
\let\auto@bib@innerbib\@empty
\bibitem [{\citenamefont {Alicea}(2012)}]{Alicea2012}%
  \BibitemOpen
  \bibfield  {author} {\bibinfo {author} {\bibfnamefont {J.}~\bibnamefont
  {Alicea}},\ }\href {\doibase 10.1088/0034-4885/75/7/076501} {\bibfield
  {journal} {\bibinfo  {journal} {Rep. Prog. Phys.}\ }\textbf {\bibinfo
  {volume} {75}},\ \bibinfo {pages} {076501} (\bibinfo {year}
  {2012})}\BibitemShut {NoStop}%
\bibitem [{\citenamefont {Beenakker}(2013)}]{Beenakker2013}%
  \BibitemOpen
  \bibfield  {author} {\bibinfo {author} {\bibfnamefont {C.~W.~J.}\
  \bibnamefont {Beenakker}},\ }\href {\doibase
  10.1146/annurev-conmatphys-030212-184337} {\bibfield  {journal} {\bibinfo
  {journal} {Ann. Rev. Condens. Matt. Phys.}\ }\textbf {\bibinfo {volume}
  {4}},\ \bibinfo {pages} {113} (\bibinfo {year} {2013})}\BibitemShut {NoStop}%
\bibitem [{\citenamefont {Kitaev}(2001)}]{Kitaev2001}%
  \BibitemOpen
  \bibfield  {author} {\bibinfo {author} {\bibfnamefont {A.~Y.}\ \bibnamefont
  {Kitaev}},\ }\href {\doibase 10.1070/1063-7869/44/10S/S29} {\bibfield
  {journal} {\bibinfo  {journal} {Phys. Usp.}\ }\textbf {\bibinfo {volume}
  {44}},\ \bibinfo {pages} {131} (\bibinfo {year} {2001})}\BibitemShut
  {NoStop}%
\bibitem [{\citenamefont {Ivanov}(2001)}]{ivanov2001non}%
  \BibitemOpen
  \bibfield  {author} {\bibinfo {author} {\bibfnamefont {D.~A.}\ \bibnamefont
  {Ivanov}},\ }\href {\doibase 10.1103/PhysRevLett.86.268} {\bibfield
  {journal} {\bibinfo  {journal} {Phys. Rev. Lett.}\ }\textbf {\bibinfo
  {volume} {86}},\ \bibinfo {pages} {268} (\bibinfo {year} {2001})}\BibitemShut
  {NoStop}%
\bibitem [{\citenamefont {Stern}\ \emph {et~al.}(2004)\citenamefont {Stern},
  \citenamefont {von Oppen},\ and\ \citenamefont
  {Mariani}}]{stern2004geometric}%
  \BibitemOpen
  \bibfield  {author} {\bibinfo {author} {\bibfnamefont {A.}~\bibnamefont
  {Stern}}, \bibinfo {author} {\bibfnamefont {F.}~\bibnamefont {von Oppen}}, \
  and\ \bibinfo {author} {\bibfnamefont {E.}~\bibnamefont {Mariani}},\ }\href
  {http://link.aps.org/doi/10.1103/PhysRevB.70.205338} {\bibfield  {journal}
  {\bibinfo  {journal} {Phys. Rev. B}\ }\textbf {\bibinfo {volume} {70}},\
  \bibinfo {pages} {205338} (\bibinfo {year} {2004})}\BibitemShut {NoStop}%
\bibitem [{\citenamefont {Nayak}\ \emph {et~al.}(2008)\citenamefont {Nayak},
  \citenamefont {Simon}, \citenamefont {Stern}, \citenamefont {Freedman},\ and\
  \citenamefont {Sarma}}]{nayak2008non}%
  \BibitemOpen
  \bibfield  {author} {\bibinfo {author} {\bibfnamefont {C.}~\bibnamefont
  {Nayak}}, \bibinfo {author} {\bibfnamefont {S.}~\bibnamefont {Simon}},
  \bibinfo {author} {\bibfnamefont {A.}~\bibnamefont {Stern}}, \bibinfo
  {author} {\bibfnamefont {M.}~\bibnamefont {Freedman}}, \ and\ \bibinfo
  {author} {\bibfnamefont {S.}~\bibnamefont {Sarma}},\ }\href
  {http://link.aps.org/doi/10.1103/RevModPhys.80.1083} {\bibfield  {journal}
  {\bibinfo  {journal} {Rev. Mod. Phys.}\ }\textbf {\bibinfo {volume} {80}},\
  \bibinfo {pages} {1083} (\bibinfo {year} {2008})}\BibitemShut {NoStop}%
\bibitem [{\citenamefont {Alicea}\ \emph {et~al.}(2011)\citenamefont {Alicea},
  \citenamefont {Oreg}, \citenamefont {Refael}, \citenamefont {von Oppen},\
  and\ \citenamefont {Fisher}}]{alicea2011non}%
  \BibitemOpen
  \bibfield  {author} {\bibinfo {author} {\bibfnamefont {J.}~\bibnamefont
  {Alicea}}, \bibinfo {author} {\bibfnamefont {Y.}~\bibnamefont {Oreg}},
  \bibinfo {author} {\bibfnamefont {G.}~\bibnamefont {Refael}}, \bibinfo
  {author} {\bibfnamefont {F.}~\bibnamefont {von Oppen}}, \ and\ \bibinfo
  {author} {\bibfnamefont {M.~P.}\ \bibnamefont {Fisher}},\ }\href
  {http://www.nature.com/nphys/journal/v7/n5/abs/nphys1915.html} {\bibfield
  {journal} {\bibinfo  {journal} {Nat. Phys.}\ }\textbf {\bibinfo {volume}
  {7}},\ \bibinfo {pages} {412} (\bibinfo {year} {2011})}\BibitemShut {NoStop}%
\bibitem [{\citenamefont {Kitaev}(2003)}]{Kitaev2003}%
  \BibitemOpen
  \bibfield  {author} {\bibinfo {author} {\bibfnamefont {A.~Y.}\ \bibnamefont
  {Kitaev}},\ }\href
  {http://www.sciencedirect.com/science/article/pii/S0003491602000180}
  {\bibfield  {journal} {\bibinfo  {journal} {Ann. Phys.}\ }\textbf {\bibinfo
  {volume} {303}},\ \bibinfo {pages} {2} (\bibinfo {year} {2003})}\BibitemShut
  {NoStop}%
\bibitem [{\citenamefont {Read}\ and\ \citenamefont
  {Green}(2000)}]{read2000paired}%
  \BibitemOpen
  \bibfield  {author} {\bibinfo {author} {\bibfnamefont {N.}~\bibnamefont
  {Read}}\ and\ \bibinfo {author} {\bibfnamefont {D.}~\bibnamefont {Green}},\
  }\href {\doibase 10.1103/PhysRevB.61.10267} {\bibfield  {journal} {\bibinfo
  {journal} {Phys. Rev. B}\ }\textbf {\bibinfo {volume} {61}},\ \bibinfo
  {pages} {10267} (\bibinfo {year} {2000})}\BibitemShut {NoStop}%
\bibitem [{\citenamefont {Moore}\ and\ \citenamefont
  {Read}(1991)}]{moore1991nonabelions}%
  \BibitemOpen
  \bibfield  {author} {\bibinfo {author} {\bibfnamefont {G.}~\bibnamefont
  {Moore}}\ and\ \bibinfo {author} {\bibfnamefont {N.}~\bibnamefont {Read}},\
  }\href {http://www.sciencedirect.com/science/article/pii/055032139190407O}
  {\bibfield  {journal} {\bibinfo  {journal} {Nucl. Phys. B}\ }\textbf
  {\bibinfo {volume} {360}},\ \bibinfo {pages} {362 } (\bibinfo {year}
  {1991})}\BibitemShut {NoStop}%
\bibitem [{\citenamefont {Fu}\ and\ \citenamefont
  {Kane}(2008)}]{fu2008superconducting}%
  \BibitemOpen
  \bibfield  {author} {\bibinfo {author} {\bibfnamefont {L.}~\bibnamefont
  {Fu}}\ and\ \bibinfo {author} {\bibfnamefont {C.~L.}\ \bibnamefont {Kane}},\
  }\href {\doibase 10.1103/PhysRevLett.100.096407} {\bibfield  {journal}
  {\bibinfo  {journal} {Phys. Rev. Lett.}\ }\textbf {\bibinfo {volume} {100}},\
  \bibinfo {pages} {096407} (\bibinfo {year} {2008})}\BibitemShut {NoStop}%
\bibitem [{\citenamefont {Fu}\ and\ \citenamefont
  {Kane}(2009)}]{fu2009josephson}%
  \BibitemOpen
  \bibfield  {author} {\bibinfo {author} {\bibfnamefont {L.}~\bibnamefont
  {Fu}}\ and\ \bibinfo {author} {\bibfnamefont {C.~L.}\ \bibnamefont {Kane}},\
  }\href {\doibase 10.1103/PhysRevB.79.161408} {\bibfield  {journal} {\bibinfo
  {journal} {Phys. Rev. B}\ }\textbf {\bibinfo {volume} {79}},\ \bibinfo
  {pages} {161408} (\bibinfo {year} {2009})}\BibitemShut {NoStop}%
\bibitem [{\citenamefont {Sau}\ \emph {et~al.}(2010)\citenamefont {Sau},
  \citenamefont {Lutchyn}, \citenamefont {Tewari},\ and\ \citenamefont
  {Das~Sarma}}]{Sau2010Generic}%
  \BibitemOpen
  \bibfield  {author} {\bibinfo {author} {\bibfnamefont {J.~D.}\ \bibnamefont
  {Sau}}, \bibinfo {author} {\bibfnamefont {R.~M.}\ \bibnamefont {Lutchyn}},
  \bibinfo {author} {\bibfnamefont {S.}~\bibnamefont {Tewari}}, \ and\ \bibinfo
  {author} {\bibfnamefont {S.}~\bibnamefont {Das~Sarma}},\ }\href {\doibase
  10.1103/PhysRevLett.104.040502} {\bibfield  {journal} {\bibinfo  {journal}
  {Phys. Rev. Lett.}\ }\textbf {\bibinfo {volume} {104}},\ \bibinfo {pages}
  {040502} (\bibinfo {year} {2010})}\BibitemShut {NoStop}%
\bibitem [{\citenamefont {Alicea}(2010)}]{Alicea2010Majorana}%
  \BibitemOpen
  \bibfield  {author} {\bibinfo {author} {\bibfnamefont {J.}~\bibnamefont
  {Alicea}},\ }\href {\doibase 10.1103/PhysRevB.81.125318} {\bibfield
  {journal} {\bibinfo  {journal} {Phys. Rev. B}\ }\textbf {\bibinfo {volume}
  {81}},\ \bibinfo {pages} {125318} (\bibinfo {year} {2010})}\BibitemShut
  {NoStop}%
\bibitem [{\citenamefont {Lutchyn}\ \emph {et~al.}(2010)\citenamefont
  {Lutchyn}, \citenamefont {Sau},\ and\ \citenamefont
  {Das~Sarma}}]{lutchyn2010majorana}%
  \BibitemOpen
  \bibfield  {author} {\bibinfo {author} {\bibfnamefont {R.~M.}\ \bibnamefont
  {Lutchyn}}, \bibinfo {author} {\bibfnamefont {J.~D.}\ \bibnamefont {Sau}}, \
  and\ \bibinfo {author} {\bibfnamefont {S.}~\bibnamefont {Das~Sarma}},\ }\href
  {\doibase 10.1103/PhysRevLett.105.077001} {\bibfield  {journal} {\bibinfo
  {journal} {Phys. Rev. Lett.}\ }\textbf {\bibinfo {volume} {105}},\ \bibinfo
  {pages} {077001} (\bibinfo {year} {2010})}\BibitemShut {NoStop}%
\bibitem [{\citenamefont {Oreg}\ \emph {et~al.}(2010)\citenamefont {Oreg},
  \citenamefont {Refael},\ and\ \citenamefont {von Oppen}}]{oreg2010helical}%
  \BibitemOpen
  \bibfield  {author} {\bibinfo {author} {\bibfnamefont {Y.}~\bibnamefont
  {Oreg}}, \bibinfo {author} {\bibfnamefont {G.}~\bibnamefont {Refael}}, \ and\
  \bibinfo {author} {\bibfnamefont {F.}~\bibnamefont {von Oppen}},\ }\href
  {\doibase 10.1103/PhysRevLett.105.177002} {\bibfield  {journal} {\bibinfo
  {journal} {Phys. Rev. Lett.}\ }\textbf {\bibinfo {volume} {105}},\ \bibinfo
  {pages} {177002} (\bibinfo {year} {2010})}\BibitemShut {NoStop}%
\bibitem [{\citenamefont {Mourik}\ \emph {et~al.}(2012)\citenamefont {Mourik},
  \citenamefont {Zuo}, \citenamefont {Frolov}, \citenamefont {Plissard},
  \citenamefont {Bakkers},\ and\ \citenamefont
  {Kouwenhoven}}]{mourik2012signatures}%
  \BibitemOpen
  \bibfield  {author} {\bibinfo {author} {\bibfnamefont {V.}~\bibnamefont
  {Mourik}}, \bibinfo {author} {\bibfnamefont {K.}~\bibnamefont {Zuo}},
  \bibinfo {author} {\bibfnamefont {S.}~\bibnamefont {Frolov}}, \bibinfo
  {author} {\bibfnamefont {S.}~\bibnamefont {Plissard}}, \bibinfo {author}
  {\bibfnamefont {E.}~\bibnamefont {Bakkers}}, \ and\ \bibinfo {author}
  {\bibfnamefont {L.}~\bibnamefont {Kouwenhoven}},\ }\href
  {http://www.sciencemag.org/content/336/6084/1003} {\bibfield  {journal}
  {\bibinfo  {journal} {Science}\ }\textbf {\bibinfo {volume} {336}},\ \bibinfo
  {pages} {1003} (\bibinfo {year} {2012})}\BibitemShut {NoStop}%
\bibitem [{\citenamefont {Deng}\ \emph {et~al.}(2012)\citenamefont {Deng},
  \citenamefont {Yu}, \citenamefont {Huang}, \citenamefont {Larsson},
  \citenamefont {Caroff},\ and\ \citenamefont {Xu}}]{Deng2012a}%
  \BibitemOpen
  \bibfield  {author} {\bibinfo {author} {\bibfnamefont {M.~T.}\ \bibnamefont
  {Deng}}, \bibinfo {author} {\bibfnamefont {C.~L.}\ \bibnamefont {Yu}},
  \bibinfo {author} {\bibfnamefont {G.~Y.}\ \bibnamefont {Huang}}, \bibinfo
  {author} {\bibfnamefont {M.}~\bibnamefont {Larsson}}, \bibinfo {author}
  {\bibfnamefont {P.}~\bibnamefont {Caroff}}, \ and\ \bibinfo {author}
  {\bibfnamefont {H.~Q.}\ \bibnamefont {Xu}},\ }\href {\doibase
  10.1021/nl303758w} {\bibfield  {journal} {\bibinfo  {journal} {Nano Lett.}\
  }\textbf {\bibinfo {volume} {12}},\ \bibinfo {pages} {6414} (\bibinfo {year}
  {2012})}\BibitemShut {NoStop}%
\bibitem [{\citenamefont {Das}\ \emph {et~al.}(2012)\citenamefont {Das},
  \citenamefont {Ronen}, \citenamefont {Most}, \citenamefont {Oreg},
  \citenamefont {Heiblum},\ and\ \citenamefont {Shtrikman}}]{Das2012zero}%
  \BibitemOpen
  \bibfield  {author} {\bibinfo {author} {\bibfnamefont {A.}~\bibnamefont
  {Das}}, \bibinfo {author} {\bibfnamefont {Y.}~\bibnamefont {Ronen}}, \bibinfo
  {author} {\bibfnamefont {Y.}~\bibnamefont {Most}}, \bibinfo {author}
  {\bibfnamefont {Y.}~\bibnamefont {Oreg}}, \bibinfo {author} {\bibfnamefont
  {M.}~\bibnamefont {Heiblum}}, \ and\ \bibinfo {author} {\bibfnamefont
  {H.}~\bibnamefont {Shtrikman}},\ }\href {\doibase 10.1038/nphys2479}
  {\bibfield  {journal} {\bibinfo  {journal} {Nat. Phys.}\ }\textbf {\bibinfo
  {volume} {8}},\ \bibinfo {pages} {887} (\bibinfo {year} {2012})}\BibitemShut
  {NoStop}%
\bibitem [{\citenamefont {Churchill}\ \emph {et~al.}(2013)\citenamefont
  {Churchill}, \citenamefont {Fatemi}, \citenamefont {Grove-Rasmussen},
  \citenamefont {Deng}, \citenamefont {Caroff}, \citenamefont {Xu},\ and\
  \citenamefont {Marcus}}]{churchill2013superconductor}%
  \BibitemOpen
  \bibfield  {author} {\bibinfo {author} {\bibfnamefont {H.~O.~H.}\
  \bibnamefont {Churchill}}, \bibinfo {author} {\bibfnamefont {V.}~\bibnamefont
  {Fatemi}}, \bibinfo {author} {\bibfnamefont {K.}~\bibnamefont
  {Grove-Rasmussen}}, \bibinfo {author} {\bibfnamefont {M.~T.}\ \bibnamefont
  {Deng}}, \bibinfo {author} {\bibfnamefont {P.}~\bibnamefont {Caroff}},
  \bibinfo {author} {\bibfnamefont {H.~Q.}\ \bibnamefont {Xu}}, \ and\ \bibinfo
  {author} {\bibfnamefont {C.~M.}\ \bibnamefont {Marcus}},\ }\href {\doibase
  10.1103/PhysRevB.87.241401} {\bibfield  {journal} {\bibinfo  {journal} {Phys.
  Rev. B}\ }\textbf {\bibinfo {volume} {87}},\ \bibinfo {pages} {241401}
  (\bibinfo {year} {2013})}\BibitemShut {NoStop}%
\bibitem [{\citenamefont {Finck}\ \emph {et~al.}(2013)\citenamefont {Finck},
  \citenamefont {Van~Harlingen}, \citenamefont {Mohseni}, \citenamefont
  {Jung},\ and\ \citenamefont {Li}}]{Finck2013}%
  \BibitemOpen
  \bibfield  {author} {\bibinfo {author} {\bibfnamefont {A.~D.~K.}\
  \bibnamefont {Finck}}, \bibinfo {author} {\bibfnamefont {D.~J.}\ \bibnamefont
  {Van~Harlingen}}, \bibinfo {author} {\bibfnamefont {P.~K.}\ \bibnamefont
  {Mohseni}}, \bibinfo {author} {\bibfnamefont {K.}~\bibnamefont {Jung}}, \
  and\ \bibinfo {author} {\bibfnamefont {X.}~\bibnamefont {Li}},\ }\href
  {\doibase 10.1103/PhysRevLett.110.126406} {\bibfield  {journal} {\bibinfo
  {journal} {Phys. Rev. Lett.}\ }\textbf {\bibinfo {volume} {110}},\ \bibinfo
  {pages} {126406} (\bibinfo {year} {2013})}\BibitemShut {NoStop}%
\bibitem [{\citenamefont {Law}\ \emph {et~al.}(2009)\citenamefont {Law},
  \citenamefont {Lee},\ and\ \citenamefont {Ng}}]{Law2009majorana}%
  \BibitemOpen
  \bibfield  {author} {\bibinfo {author} {\bibfnamefont {K.~T.}\ \bibnamefont
  {Law}}, \bibinfo {author} {\bibfnamefont {P.~A.}\ \bibnamefont {Lee}}, \ and\
  \bibinfo {author} {\bibfnamefont {T.~K.}\ \bibnamefont {Ng}},\ }\href
  {\doibase 10.1103/PhysRevLett.103.237001} {\bibfield  {journal} {\bibinfo
  {journal} {Phys. Rev. Lett.}\ }\textbf {\bibinfo {volume} {103}},\ \bibinfo
  {pages} {237001} (\bibinfo {year} {2009})}\BibitemShut {NoStop}%
\bibitem [{\citenamefont {Bolech}\ and\ \citenamefont
  {Demler}(2007)}]{Bolech2007Observing}%
  \BibitemOpen
  \bibfield  {author} {\bibinfo {author} {\bibfnamefont {C.~J.}\ \bibnamefont
  {Bolech}}\ and\ \bibinfo {author} {\bibfnamefont {E.}~\bibnamefont
  {Demler}},\ }\href {\doibase 10.1103/PhysRevLett.98.237002} {\bibfield
  {journal} {\bibinfo  {journal} {Phys. Rev. Lett.}\ }\textbf {\bibinfo
  {volume} {98}},\ \bibinfo {pages} {237002} (\bibinfo {year}
  {2007})}\BibitemShut {NoStop}%
\bibitem [{\citenamefont {Fidkowski}\ \emph {et~al.}(2012)\citenamefont
  {Fidkowski}, \citenamefont {Alicea}, \citenamefont {Lindner}, \citenamefont
  {Lutchyn},\ and\ \citenamefont {Fisher}}]{Fidkowski2012universal}%
  \BibitemOpen
  \bibfield  {author} {\bibinfo {author} {\bibfnamefont {L.}~\bibnamefont
  {Fidkowski}}, \bibinfo {author} {\bibfnamefont {J.}~\bibnamefont {Alicea}},
  \bibinfo {author} {\bibfnamefont {N.~H.}\ \bibnamefont {Lindner}}, \bibinfo
  {author} {\bibfnamefont {R.~M.}\ \bibnamefont {Lutchyn}}, \ and\ \bibinfo
  {author} {\bibfnamefont {M.~P.~A.}\ \bibnamefont {Fisher}},\ }\href {\doibase
  10.1103/PhysRevB.85.245121} {\bibfield  {journal} {\bibinfo  {journal} {Phys.
  Rev. B}\ }\textbf {\bibinfo {volume} {85}},\ \bibinfo {pages} {245121}
  (\bibinfo {year} {2012})}\BibitemShut {NoStop}%
\bibitem [{\citenamefont {Nadj-Perge}\ \emph {et~al.}(2014)\citenamefont
  {Nadj-Perge}, \citenamefont {Drozdov}, \citenamefont {Li}, \citenamefont
  {Chen}, \citenamefont {Jeon}, \citenamefont {Seo}, \citenamefont {MacDonald},
  \citenamefont {Bernevig},\ and\ \citenamefont
  {Yazdani}}]{Nadj-Perge2014observation}%
  \BibitemOpen
  \bibfield  {author} {\bibinfo {author} {\bibfnamefont {S.}~\bibnamefont
  {Nadj-Perge}}, \bibinfo {author} {\bibfnamefont {I.~K.}\ \bibnamefont
  {Drozdov}}, \bibinfo {author} {\bibfnamefont {J.}~\bibnamefont {Li}},
  \bibinfo {author} {\bibfnamefont {H.}~\bibnamefont {Chen}}, \bibinfo {author}
  {\bibfnamefont {S.}~\bibnamefont {Jeon}}, \bibinfo {author} {\bibfnamefont
  {J.}~\bibnamefont {Seo}}, \bibinfo {author} {\bibfnamefont {A.~H.}\
  \bibnamefont {MacDonald}}, \bibinfo {author} {\bibfnamefont {B.~A.}\
  \bibnamefont {Bernevig}}, \ and\ \bibinfo {author} {\bibfnamefont
  {A.}~\bibnamefont {Yazdani}},\ }\href
  {http://www.sciencemag.org/content/early/2014/10/01/science.1259327.abstract}
  {\bibfield  {journal} {\bibinfo  {journal} {Science}\ }\textbf {\bibinfo
  {volume} {346}},\ \bibinfo {pages} {602} (\bibinfo {year}
  {2014})}\BibitemShut {NoStop}%
\bibitem [{\citenamefont {Pikulin}\ \emph {et~al.}(2012)\citenamefont
  {Pikulin}, \citenamefont {Dahlhaus}, \citenamefont {Wimmer}, \citenamefont
  {Schomerus},\ and\ \citenamefont {Beenakker}}]{Pikulin2012}%
  \BibitemOpen
  \bibfield  {author} {\bibinfo {author} {\bibfnamefont {D.~I.}\ \bibnamefont
  {Pikulin}}, \bibinfo {author} {\bibfnamefont {J.~P.}\ \bibnamefont
  {Dahlhaus}}, \bibinfo {author} {\bibfnamefont {M.}~\bibnamefont {Wimmer}},
  \bibinfo {author} {\bibfnamefont {H.}~\bibnamefont {Schomerus}}, \ and\
  \bibinfo {author} {\bibfnamefont {C.~W.~J.}\ \bibnamefont {Beenakker}},\
  }\href {http://stacks.iop.org/1367-2630/14/i=12/a=125011} {\bibfield
  {journal} {\bibinfo  {journal} {New Journal of Physics}\ }\textbf {\bibinfo
  {volume} {14}},\ \bibinfo {pages} {125011} (\bibinfo {year}
  {2012})}\BibitemShut {NoStop}%
\bibitem [{\citenamefont {Kells}\ \emph {et~al.}(2012)\citenamefont {Kells},
  \citenamefont {Meidan},\ and\ \citenamefont {Brouwer}}]{kells2012near}%
  \BibitemOpen
  \bibfield  {author} {\bibinfo {author} {\bibfnamefont {G.}~\bibnamefont
  {Kells}}, \bibinfo {author} {\bibfnamefont {D.}~\bibnamefont {Meidan}}, \
  and\ \bibinfo {author} {\bibfnamefont {P.~W.}\ \bibnamefont {Brouwer}},\
  }\href {\doibase 10.1103/PhysRevB.86.100503} {\bibfield  {journal} {\bibinfo
  {journal} {Phys. Rev. B}\ }\textbf {\bibinfo {volume} {86}},\ \bibinfo
  {pages} {100503} (\bibinfo {year} {2012})}\BibitemShut {NoStop}%
\bibitem [{\citenamefont {Liu}\ \emph {et~al.}(2012)\citenamefont {Liu},
  \citenamefont {Potter}, \citenamefont {Law},\ and\ \citenamefont
  {Lee}}]{liu2012zero}%
  \BibitemOpen
  \bibfield  {author} {\bibinfo {author} {\bibfnamefont {J.}~\bibnamefont
  {Liu}}, \bibinfo {author} {\bibfnamefont {A.~C.}\ \bibnamefont {Potter}},
  \bibinfo {author} {\bibfnamefont {K.~T.}\ \bibnamefont {Law}}, \ and\
  \bibinfo {author} {\bibfnamefont {P.~A.}\ \bibnamefont {Lee}},\ }\href
  {\doibase 10.1103/PhysRevLett.109.267002} {\bibfield  {journal} {\bibinfo
  {journal} {Phys. Rev. Lett.}\ }\textbf {\bibinfo {volume} {109}},\ \bibinfo
  {pages} {267002} (\bibinfo {year} {2012})}\BibitemShut {NoStop}%
\bibitem [{\citenamefont {Sasaki}\ \emph {et~al.}(2000)\citenamefont {Sasaki},
  \citenamefont {De~Franceschi}, \citenamefont {Elzerman}, \citenamefont
  {Van~der Wiel}, \citenamefont {Eto}, \citenamefont {Tarucha},\ and\
  \citenamefont {Kouwenhoven}}]{sasaki2000kondo}%
  \BibitemOpen
  \bibfield  {author} {\bibinfo {author} {\bibfnamefont {S.}~\bibnamefont
  {Sasaki}}, \bibinfo {author} {\bibfnamefont {S.}~\bibnamefont
  {De~Franceschi}}, \bibinfo {author} {\bibfnamefont {J.}~\bibnamefont
  {Elzerman}}, \bibinfo {author} {\bibfnamefont {W.}~\bibnamefont {Van~der
  Wiel}}, \bibinfo {author} {\bibfnamefont {M.}~\bibnamefont {Eto}}, \bibinfo
  {author} {\bibfnamefont {S.}~\bibnamefont {Tarucha}}, \ and\ \bibinfo
  {author} {\bibfnamefont {L.}~\bibnamefont {Kouwenhoven}},\ }\href
  {http://www.nature.com/nature/journal/v405/n6788/abs/405764a0.html}
  {\bibfield  {journal} {\bibinfo  {journal} {Nature (London)}\ }\textbf
  {\bibinfo {volume} {405}},\ \bibinfo {pages} {764} (\bibinfo {year}
  {2000})}\BibitemShut {NoStop}%
\bibitem [{\citenamefont {Bagrets}\ and\ \citenamefont
  {Altland}(2012)}]{Bagrets2012Class}%
  \BibitemOpen
  \bibfield  {author} {\bibinfo {author} {\bibfnamefont {D.}~\bibnamefont
  {Bagrets}}\ and\ \bibinfo {author} {\bibfnamefont {A.}~\bibnamefont
  {Altland}},\ }\href {\doibase 10.1103/PhysRevLett.109.227005} {\bibfield
  {journal} {\bibinfo  {journal} {Phys. Rev. Lett.}\ }\textbf {\bibinfo
  {volume} {109}},\ \bibinfo {pages} {227005} (\bibinfo {year}
  {2012})}\BibitemShut {NoStop}%
\bibitem [{\citenamefont {Lee}\ \emph {et~al.}(2014)\citenamefont {Lee},
  \citenamefont {Jiang}, \citenamefont {Houzet}, \citenamefont {Aguado},
  \citenamefont {Lieber},\ and\ \citenamefont {De~Franceschi}}]{lee2013spin}%
  \BibitemOpen
  \bibfield  {author} {\bibinfo {author} {\bibfnamefont {E.~J.}\ \bibnamefont
  {Lee}}, \bibinfo {author} {\bibfnamefont {X.}~\bibnamefont {Jiang}}, \bibinfo
  {author} {\bibfnamefont {M.}~\bibnamefont {Houzet}}, \bibinfo {author}
  {\bibfnamefont {R.}~\bibnamefont {Aguado}}, \bibinfo {author} {\bibfnamefont
  {C.~M.}\ \bibnamefont {Lieber}}, \ and\ \bibinfo {author} {\bibfnamefont
  {S.}~\bibnamefont {De~Franceschi}},\ }\href
  {http://www.nature.com/nnano/journal/v9/n1/full/nnano.2013.267.html}
  {\bibfield  {journal} {\bibinfo  {journal} {Nat. nanotechnology}\ }\textbf
  {\bibinfo {volume} {9}},\ \bibinfo {pages} {79} (\bibinfo {year}
  {2014})}\BibitemShut {NoStop}%
\bibitem [{\citenamefont {Andreev}(1964)}]{andreev1964thermal}%
  \BibitemOpen
  \bibfield  {author} {\bibinfo {author} {\bibfnamefont {A.}~\bibnamefont
  {Andreev}},\ }\href
  {http://www.jetp.ac.ru/cgi-bin/e/index/r/46/5/p1823?a=list} {\bibfield
  {journal} {\bibinfo  {journal} {Zh. Eksp. Teor. Fiz.}\ }\textbf {\bibinfo
  {volume} {46}},\ \bibinfo {pages} {1823} (\bibinfo {year}
  {1964})}\BibitemShut {NoStop}%
\bibitem [{cha()}]{changing_BS_occu}%
  \BibitemOpen
  \href@noop {} {}\bibinfo {note} {Note that this process can happen more than
  once since charge is not conserved in the superconductor.}\BibitemShut
  {Stop}%
\bibitem [{equ()}]{equal_spin_density}%
  \BibitemOpen
  \href@noop {} {}\bibinfo {note} {Formally this reads $\langle0|{\bf
  s}(x)|0\rangle=\langle1|{\bf s}(x)|1\rangle$, where ${\bf s}(x)$ is the spin
  density and $|0\rangle, |1\rangle$ are the two degenerate ground
  states.}\BibitemShut {Stop}%
\bibitem [{mul()}]{multi_channel_2}%
  \BibitemOpen
  \href@noop {} {}\bibinfo {note} {Although we consider here a single-channel
  lead, our conclusions extend to a multichannel lead.}\BibitemShut {Stop}%
\bibitem [{\citenamefont {Fisher}\ and\ \citenamefont
  {Lee}(1981)}]{Fisher1981relation}%
  \BibitemOpen
  \bibfield  {author} {\bibinfo {author} {\bibfnamefont {D.~S.}\ \bibnamefont
  {Fisher}}\ and\ \bibinfo {author} {\bibfnamefont {P.~A.}\ \bibnamefont
  {Lee}},\ }\href {\doibase 10.1103/PhysRevB.23.6851} {\bibfield  {journal}
  {\bibinfo  {journal} {Phys. Rev. B}\ }\textbf {\bibinfo {volume} {23}},\
  \bibinfo {pages} {6851} (\bibinfo {year} {1981})}\BibitemShut {NoStop}%
\bibitem [{\citenamefont {Iida}\ \emph {et~al.}(1990)\citenamefont {Iida},
  \citenamefont {Weidenmüller},\ and\ \citenamefont
  {Zuk}}]{Iida1990statistical}%
  \BibitemOpen
  \bibfield  {author} {\bibinfo {author} {\bibfnamefont {S.}~\bibnamefont
  {Iida}}, \bibinfo {author} {\bibfnamefont {H.}~\bibnamefont {Weidenmüller}},
  \ and\ \bibinfo {author} {\bibfnamefont {J.}~\bibnamefont {Zuk}},\ }\href
  {http://www.sciencedirect.com/science/article/pii/000349169090275S}
  {\bibfield  {journal} {\bibinfo  {journal} {Ann. Phys.}\ }\textbf {\bibinfo
  {volume} {200}},\ \bibinfo {pages} {219 } (\bibinfo {year}
  {1990})}\BibitemShut {NoStop}%
\bibitem [{\citenamefont {Anantram}\ and\ \citenamefont
  {Datta}(1996)}]{Anantram1996current}%
  \BibitemOpen
  \bibfield  {author} {\bibinfo {author} {\bibfnamefont {M.~P.}\ \bibnamefont
  {Anantram}}\ and\ \bibinfo {author} {\bibfnamefont {S.}~\bibnamefont
  {Datta}},\ }\href {\doibase 10.1103/PhysRevB.53.16390} {\bibfield  {journal}
  {\bibinfo  {journal} {Phys. Rev. B}\ }\textbf {\bibinfo {volume} {53}},\
  \bibinfo {pages} {16390} (\bibinfo {year} {1996})}\BibitemShut {NoStop}%
\bibitem [{\citenamefont {Golub}\ and\ \citenamefont
  {Horovitz}(2011)}]{Golub2011shot}%
  \BibitemOpen
  \bibfield  {author} {\bibinfo {author} {\bibfnamefont {A.}~\bibnamefont
  {Golub}}\ and\ \bibinfo {author} {\bibfnamefont {B.}~\bibnamefont
  {Horovitz}},\ }\href {\doibase 10.1103/PhysRevB.83.153415} {\bibfield
  {journal} {\bibinfo  {journal} {Phys. Rev. B}\ }\textbf {\bibinfo {volume}
  {83}},\ \bibinfo {pages} {153415} (\bibinfo {year} {2011})}\BibitemShut
  {NoStop}%
\bibitem [{\citenamefont {Golub}\ and\ \citenamefont
  {Horovitz}()}]{golub2014interferometric}%
  \BibitemOpen
  \bibfield  {author} {\bibinfo {author} {\bibfnamefont {A.}~\bibnamefont
  {Golub}}\ and\ \bibinfo {author} {\bibfnamefont {B.}~\bibnamefont
  {Horovitz}},\ }\href {http://arxiv.org/abs/1407.5179} {\bibinfo  {journal}
  {arXiv:1407.5179}\ }\BibitemShut {NoStop}%
\bibitem [{\citenamefont {Nilsson}\ \emph {et~al.}(2008)\citenamefont
  {Nilsson}, \citenamefont {Akhmerov},\ and\ \citenamefont
  {Beenakker}}]{Nilsson2008splitting}%
  \BibitemOpen
\bibfield  {journal} {  }\bibfield  {author} {\bibinfo {author} {\bibfnamefont
  {J.}~\bibnamefont {Nilsson}}, \bibinfo {author} {\bibfnamefont {A.~R.}\
  \bibnamefont {Akhmerov}}, \ and\ \bibinfo {author} {\bibfnamefont {C.~W.~J.}\
  \bibnamefont {Beenakker}},\ }\href {\doibase 10.1103/PhysRevLett.101.120403}
  {\bibfield  {journal} {\bibinfo  {journal} {Phys. Rev. Lett.}\ }\textbf
  {\bibinfo {volume} {101}},\ \bibinfo {pages} {120403} (\bibinfo {year}
  {2008})}\BibitemShut {NoStop}%
\bibitem [{\citenamefont {Leijnse}\ and\ \citenamefont
  {Flensberg}(2011)}]{Leijnse2011quantum}%
  \BibitemOpen
  \bibfield  {author} {\bibinfo {author} {\bibfnamefont {M.}~\bibnamefont
  {Leijnse}}\ and\ \bibinfo {author} {\bibfnamefont {K.}~\bibnamefont
  {Flensberg}},\ }\href {\doibase 10.1103/PhysRevLett.107.210502} {\bibfield
  {journal} {\bibinfo  {journal} {Phys. Rev. Lett.}\ }\textbf {\bibinfo
  {volume} {107}},\ \bibinfo {pages} {210502} (\bibinfo {year}
  {2011})}\BibitemShut {NoStop}%
\bibitem [{\citenamefont {He}\ \emph {et~al.}(2014)\citenamefont {He},
  \citenamefont {Ng}, \citenamefont {Lee},\ and\ \citenamefont
  {Law}}]{He2014selective}%
  \BibitemOpen
  \bibfield  {author} {\bibinfo {author} {\bibfnamefont {J.~J.}\ \bibnamefont
  {He}}, \bibinfo {author} {\bibfnamefont {T.~K.}\ \bibnamefont {Ng}}, \bibinfo
  {author} {\bibfnamefont {P.~A.}\ \bibnamefont {Lee}}, \ and\ \bibinfo
  {author} {\bibfnamefont {K.~T.}\ \bibnamefont {Law}},\ }\href {\doibase
  10.1103/PhysRevLett.112.037001} {\bibfield  {journal} {\bibinfo  {journal}
  {Phys. Rev. Lett.}\ }\textbf {\bibinfo {volume} {112}},\ \bibinfo {pages}
  {037001} (\bibinfo {year} {2014})}\BibitemShut {NoStop}%
\bibitem [{Foo()}]{Footnote}%
  \BibitemOpen
  \href@noop {} {}\bibinfo {note} {We shall show using numerical simulations of
  a system with spin-orbit coupling that our conclusions apply also for an ABS
  with a non-well-defined spin.}\BibitemShut {Stop}%
\bibitem [{SM()}]{SM}%
  \BibitemOpen
  \href@noop {} {}\bibinfo {note} {See Supplemental Material.}\BibitemShut
  {Stop}%
\bibitem [{\citenamefont {Lim}\ \emph {et~al.}(2012)\citenamefont {Lim},
  \citenamefont {Serra}, \citenamefont {L\'opez},\ and\ \citenamefont
  {Aguado}}]{Lim2012magnetic}%
  \BibitemOpen
  \bibfield  {author} {\bibinfo {author} {\bibfnamefont {J.~S.}\ \bibnamefont
  {Lim}}, \bibinfo {author} {\bibfnamefont {L.}~\bibnamefont {Serra}}, \bibinfo
  {author} {\bibfnamefont {R.}~\bibnamefont {L\'opez}}, \ and\ \bibinfo
  {author} {\bibfnamefont {R.}~\bibnamefont {Aguado}},\ }\href {\doibase
  10.1103/PhysRevB.86.121103} {\bibfield  {journal} {\bibinfo  {journal} {Phys.
  Rev. B}\ }\textbf {\bibinfo {volume} {86}},\ \bibinfo {pages} {121103}
  (\bibinfo {year} {2012})}\BibitemShut {NoStop}%
\bibitem [{\citenamefont {Das~Sarma}\ \emph {et~al.}(2012)\citenamefont
  {Das~Sarma}, \citenamefont {Sau},\ and\ \citenamefont
  {Stanescu}}]{Das-Sarma2012splitting}%
  \BibitemOpen
  \bibfield  {author} {\bibinfo {author} {\bibfnamefont {S.}~\bibnamefont
  {Das~Sarma}}, \bibinfo {author} {\bibfnamefont {J.~D.}\ \bibnamefont {Sau}},
  \ and\ \bibinfo {author} {\bibfnamefont {T.~D.}\ \bibnamefont {Stanescu}},\
  }\href {http://link.aps.org/doi/10.1103/PhysRevB.86.220506} {\bibfield
  {journal} {\bibinfo  {journal} {Phys. Rev. B}\ }\textbf {\bibinfo {volume}
  {86}},\ \bibinfo {pages} {220506} (\bibinfo {year} {2012})}\BibitemShut
  {NoStop}%
\bibitem [{\citenamefont {Rainis}\ \emph {et~al.}(2013)\citenamefont {Rainis},
  \citenamefont {Trifunovic}, \citenamefont {Klinovaja},\ and\ \citenamefont
  {Loss}}]{Rainis2013Towards}%
  \BibitemOpen
  \bibfield  {author} {\bibinfo {author} {\bibfnamefont {D.}~\bibnamefont
  {Rainis}}, \bibinfo {author} {\bibfnamefont {L.}~\bibnamefont {Trifunovic}},
  \bibinfo {author} {\bibfnamefont {J.}~\bibnamefont {Klinovaja}}, \ and\
  \bibinfo {author} {\bibfnamefont {D.}~\bibnamefont {Loss}},\ }\href {\doibase
  10.1103/PhysRevB.87.024515} {\bibfield  {journal} {\bibinfo  {journal} {Phys.
  Rev. B}\ }\textbf {\bibinfo {volume} {87}},\ \bibinfo {pages} {024515}
  (\bibinfo {year} {2013})}\BibitemShut {NoStop}%
\bibitem [{Kli()}]{Klinovaja2012Composite}%
  \BibitemOpen
  \href@noop {} {}\bibinfo {note} {Note that the wave function has support in
  the uncovered part of the wire, see J. Klinovaja and D. Loss,
  \href{http://link.aps.org/doi/10.1103/PhysRevB.86.085408}{Phys. Rev. B {\bf
  86}, 085408 (2012)}}\BibitemShut {NoStop}%
\bibitem [{\citenamefont {Sticlet}\ \emph {et~al.}(2012)\citenamefont
  {Sticlet}, \citenamefont {Bena},\ and\ \citenamefont
  {Simon}}]{Sticlet2012spin}%
  \BibitemOpen
  \bibfield  {author} {\bibinfo {author} {\bibfnamefont {D.}~\bibnamefont
  {Sticlet}}, \bibinfo {author} {\bibfnamefont {C.}~\bibnamefont {Bena}}, \
  and\ \bibinfo {author} {\bibfnamefont {P.}~\bibnamefont {Simon}},\ }\href
  {\doibase 10.1103/PhysRevLett.108.096802} {\bibfield  {journal} {\bibinfo
  {journal} {Phys. Rev. Lett.}\ }\textbf {\bibinfo {volume} {108}},\ \bibinfo
  {pages} {096802} (\bibinfo {year} {2012})}\BibitemShut {NoStop}%
\bibitem [{\citenamefont {Guigou}\ \emph {et~al.}()\citenamefont {Guigou},
  \citenamefont {Sedlmayr}, \citenamefont {Aguiar-Hualde},\ and\ \citenamefont
  {Bena}}]{guigou2014signature}%
  \BibitemOpen
  \bibfield  {author} {\bibinfo {author} {\bibfnamefont {M.}~\bibnamefont
  {Guigou}}, \bibinfo {author} {\bibfnamefont {N.}~\bibnamefont {Sedlmayr}},
  \bibinfo {author} {\bibfnamefont {J.}~\bibnamefont {Aguiar-Hualde}}, \ and\
  \bibinfo {author} {\bibfnamefont {C.}~\bibnamefont {Bena}},\ }\href
  {http://arxiv.org/abs/1407.1393} {\bibinfo  {journal} {arXiv:1407.1393}\
  }\BibitemShut {NoStop}%
\bibitem [{\citenamefont {Cuevas}\ \emph {et~al.}(2001)\citenamefont {Cuevas},
  \citenamefont {Levy~Yeyati},\ and\ \citenamefont
  {Mart\'in-Rodero}}]{Cuevas2001kondo}%
  \BibitemOpen
\bibfield  {journal} {  }\bibfield  {author} {\bibinfo {author} {\bibfnamefont
  {J.~C.}\ \bibnamefont {Cuevas}}, \bibinfo {author} {\bibfnamefont
  {A.}~\bibnamefont {Levy~Yeyati}}, \ and\ \bibinfo {author} {\bibfnamefont
  {A.}~\bibnamefont {Mart\'in-Rodero}},\ }\href {\doibase
  10.1103/PhysRevB.63.094515} {\bibfield  {journal} {\bibinfo  {journal} {Phys.
  Rev. B}\ }\textbf {\bibinfo {volume} {63}},\ \bibinfo {pages} {094515}
  (\bibinfo {year} {2001})}\BibitemShut {NoStop}%
\bibitem [{\citenamefont {Lee}\ \emph {et~al.}(2012)\citenamefont {Lee},
  \citenamefont {Jiang}, \citenamefont {Aguado}, \citenamefont {Katsaros},
  \citenamefont {Lieber},\ and\ \citenamefont
  {De~Franceschi}}]{Lee2012zero-bias}%
  \BibitemOpen
  \bibfield  {author} {\bibinfo {author} {\bibfnamefont {E.~J.~H.}\
  \bibnamefont {Lee}}, \bibinfo {author} {\bibfnamefont {X.}~\bibnamefont
  {Jiang}}, \bibinfo {author} {\bibfnamefont {R.}~\bibnamefont {Aguado}},
  \bibinfo {author} {\bibfnamefont {G.}~\bibnamefont {Katsaros}}, \bibinfo
  {author} {\bibfnamefont {C.~M.}\ \bibnamefont {Lieber}}, \ and\ \bibinfo
  {author} {\bibfnamefont {S.}~\bibnamefont {De~Franceschi}},\ }\href {\doibase
  10.1103/PhysRevLett.109.186802} {\bibfield  {journal} {\bibinfo  {journal}
  {Phys. Rev. Lett.}\ }\textbf {\bibinfo {volume} {109}},\ \bibinfo {pages}
  {186802} (\bibinfo {year} {2012})}\BibitemShut {NoStop}%
\bibitem [{\citenamefont {Cheng}\ \emph {et~al.}(2014)\citenamefont {Cheng},
  \citenamefont {Becker}, \citenamefont {Bauer},\ and\ \citenamefont
  {Lutchyn}}]{Cheng2014interplay}%
  \BibitemOpen
  \bibfield  {author} {\bibinfo {author} {\bibfnamefont {M.}~\bibnamefont
  {Cheng}}, \bibinfo {author} {\bibfnamefont {M.}~\bibnamefont {Becker}},
  \bibinfo {author} {\bibfnamefont {B.}~\bibnamefont {Bauer}}, \ and\ \bibinfo
  {author} {\bibfnamefont {R.~M.}\ \bibnamefont {Lutchyn}},\ }\href {\doibase
  10.1103/PhysRevX.4.031051} {\bibfield  {journal} {\bibinfo  {journal} {Phys.
  Rev. X}\ }\textbf {\bibinfo {volume} {4}},\ \bibinfo {pages} {031051}
  (\bibinfo {year} {2014})}\BibitemShut {NoStop}%
\bibitem [{FN_()}]{FN_2}%
  \BibitemOpen
  \href@noop {} {}\bibinfo {note} {The width of the resonance has to be
  comparable with the voltage range of interest. Since $B$ is commonly larger
  than the excitation gap, this can be achieved without allowing transmission
  of opposite spins.}\BibitemShut {Stop}%
\bibitem [{\citenamefont {Posske}\ and\ \citenamefont
  {Trauzettel}(2014)}]{Posske2014direct}%
  \BibitemOpen
  \bibfield  {author} {\bibinfo {author} {\bibfnamefont {T.}~\bibnamefont
  {Posske}}\ and\ \bibinfo {author} {\bibfnamefont {B.}~\bibnamefont
  {Trauzettel}},\ }\href {\doibase 10.1103/PhysRevB.89.075108} {\bibfield
  {journal} {\bibinfo  {journal} {Phys. Rev. B}\ }\textbf {\bibinfo {volume}
  {89}},\ \bibinfo {pages} {075108} (\bibinfo {year} {2014})}\BibitemShut
  {NoStop}%
\bibitem [{\citenamefont {Kane}\ and\ \citenamefont
  {Mele}(2005)}]{Kane2005quantum}%
  \BibitemOpen
  \bibfield  {author} {\bibinfo {author} {\bibfnamefont {C.~L.}\ \bibnamefont
  {Kane}}\ and\ \bibinfo {author} {\bibfnamefont {E.~J.}\ \bibnamefont
  {Mele}},\ }\href {\doibase 10.1103/PhysRevLett.95.226801} {\bibfield
  {journal} {\bibinfo  {journal} {Phys. Rev. Lett.}\ }\textbf {\bibinfo
  {volume} {95}},\ \bibinfo {pages} {226801} (\bibinfo {year}
  {2005})}\BibitemShut {NoStop}%
\bibitem [{\citenamefont {Bernevig}\ \emph {et~al.}(2006)\citenamefont
  {Bernevig}, \citenamefont {Hughes},\ and\ \citenamefont
  {Zhang}}]{Bernevig2006quantum}%
  \BibitemOpen
  \bibfield  {author} {\bibinfo {author} {\bibfnamefont {B.~A.}\ \bibnamefont
  {Bernevig}}, \bibinfo {author} {\bibfnamefont {T.~L.}\ \bibnamefont
  {Hughes}}, \ and\ \bibinfo {author} {\bibfnamefont {S.-C.}\ \bibnamefont
  {Zhang}},\ }\href {\doibase 10.1126/science.1133734} {\bibfield  {journal}
  {\bibinfo  {journal} {Science}\ }\textbf {\bibinfo {volume} {314}},\ \bibinfo
  {pages} {1757} (\bibinfo {year} {2006})}\BibitemShut {NoStop}%
\bibitem [{\citenamefont {K{\"o}nig}\ \emph {et~al.}(2007)\citenamefont
  {K{\"o}nig}, \citenamefont {Wiedmann}, \citenamefont {Br{\"u}ne},
  \citenamefont {Roth}, \citenamefont {Buhmann}, \citenamefont {Molenkamp},
  \citenamefont {Qi},\ and\ \citenamefont {Zhang}}]{konig2007quantum}%
  \BibitemOpen
  \bibfield  {author} {\bibinfo {author} {\bibfnamefont {M.}~\bibnamefont
  {K{\"o}nig}}, \bibinfo {author} {\bibfnamefont {S.}~\bibnamefont {Wiedmann}},
  \bibinfo {author} {\bibfnamefont {C.}~\bibnamefont {Br{\"u}ne}}, \bibinfo
  {author} {\bibfnamefont {A.}~\bibnamefont {Roth}}, \bibinfo {author}
  {\bibfnamefont {H.}~\bibnamefont {Buhmann}}, \bibinfo {author} {\bibfnamefont
  {L.~W.}\ \bibnamefont {Molenkamp}}, \bibinfo {author} {\bibfnamefont {X.-L.}\
  \bibnamefont {Qi}}, \ and\ \bibinfo {author} {\bibfnamefont {S.-C.}\
  \bibnamefont {Zhang}},\ }\href
  {http://www.sciencemag.org/content/318/5851/766.short} {\bibfield  {journal}
  {\bibinfo  {journal} {Science}\ }\textbf {\bibinfo {volume} {318}},\ \bibinfo
  {pages} {766} (\bibinfo {year} {2007})}\BibitemShut {NoStop}%
\end{thebibliography}%

\newpage

\section{{\fontsize{12}{1em}\selectfont Supplementary material}}

\section{Tight-binding simulation}
\label{TB_sim}
To obtain the scattering matrix for the system depicted in Fig.~$1$ of the main text we discretize the Hamiltonian of Eq.~(12) of the main text on a 1D lattice of $N$ sites, resulting in the following tight-binding Hamiltonian:
\begin{equation}
\begin{split}
H=&\sum_{i=1}^{N}\sum_{s,s'}[(-\mu+2t)\delta_{ss'}+\mathbf{B}\cdot\boldsymbol{\sigma}_{ss'}]c_{i,s}^\dag c_{i,s'}\\ -&[(t\delta_{ss'}-iu\sigma^z_{ss'})c_{i,s}^\dag c_{i+1,s'}+{\rm H.c.}]\\
+& \sum_{i=1}^{N}[\Delta_i c_{i,\uparrow}^\dag c^\dag_{i,\downarrow}+{\rm H.c.}]=\displaystyle\sum_{m,n=1}^{4N}\Psi^\dag_m\mathcal{H}^{\rm BdG}_{mn}\Psi_n
\end{split}\label{eq:TB_Hamiltonian}
\end{equation}
where $\Psi^\dag=(\begin{matrix}c^\dag_\mathsmaller{1\uparrow},&c^\dag_\mathsmaller{1\downarrow},...&c^\dag_\mathsmaller{N\uparrow},& c^\dag_\mathsmaller{N\downarrow}, c_\mathsmaller{1\uparrow},&c_\mathsmaller{1\downarrow},...&c_\mathsmaller{N\uparrow},&c_\mathsmaller{N\downarrow}\end{matrix}$), and
\begin{equation}
\Delta_i=\left\{
\begin{array}{lcc}
1\le i\le (N-N_S)/2, & 0\\
(N-N_S)/2< i\le (N+N_S)/2, & \Delta_0\\
(N+N_S)/2< i\le N, & 0
\end{array}
\right.
\end{equation}

We wish to relate the tight-binding parameters $t$, $u$, $N$ and $N_{\rm S}$ in Eq.~\eqref{eq:TB_Hamiltonian} to the physical parameters which appear in the main text. To this end we first note that the spin-orbit coupling energy and spin-orbit length are given by $E_{\rm so}=u^2/t$ and $l_{\rm so}=ta/u$ respectively, with $a=L/N$ being the lattice spacing. To adequately describe a continuous wire using a tight-binding model we require that the bandwidth $4t$ is much larger than all other energy scales. In the present work we take $4t=40\Delta_0$. Given $E_{\rm so}$, $l_{\rm so}$, and the wire length $L$, the parameters $u$ and $N$ are therefore determined. Finally we have $N_{\rm S}=N\cdot L_{\rm S}/L$. For the system parameters of the present work this resulted in $u=1.4\Delta_0$, $N=90$ and $N_{\rm S}=40$. The chemical potential used in the simulations is $\mu=0.5\Delta_0=125\mu eV$.

\subsection{Scattering matrix}

Using numerical matrix inversion we obtain the $4\times4$ scattering matrix of the system with the help of~\cite{Fisher1981relation,Iida1990statistical}
\begin{equation}
S(E) = 1-2\pi iW^\dag\left(E\cdot\mathbb{1}-\mathcal{H}^{\rm BdG}+i\pi W W^\dag\right)^{-1}W\label{eq:Weidenmuller},
\end{equation}
where $W$ is a $4N\times4$ matrix given by
\begin{equation}
W_{i,j}=w_0\cdot\left\{
\begin{array}{rcr}
1&, & i=1,j=1\\
1&, & i=2,j=2\\
1&, & i=2N+1,j=3\\
1&, & i=2N+2,j=4\\
0&, & {\rm OW.}
\end{array}
\right. ,
\end{equation}
and $S(E)$ contains four $2\times2$ reflection blocks
\begin{equation}
S(E)=\begin{pmatrix}r^{ee}&r^{eh}\\r^{he}&r^{hh}\end{pmatrix}.
\end{equation}
In the simulations described in the main text we have used $w^2_0=0.25\Delta_0$.

\subsection{Local density of states}
The local density of states presented in Fig.~$2$b of the main text for the system of a wire decoupled from a lead is given in terms of the Green's function
\begin{equation}
\mathcal{N}(x_i,E)=-\frac{1}{\pi}\Imag \sum_{\renewcommand{\arraystretch}{0.3}\begin{array}{c}\mathsmaller{\alpha\in e,h}\\\mathsmaller{s\in\uparrow,\downarrow}\end{array}} [G^{\rm R}(E)]^{\alpha\alpha}_{ii;ss} ~,
\end{equation}
where $G^{\rm R}(E)$ is a $4N\times4N$ matrix obtained by numerically inverting the BdG Hamiltonian:
\begin{equation}
G^{\rm R}(E) = \left[E\cdot\mathbb{1}-\mathcal{H}^{\rm BdG}+i\eta\cdot\mathbb{1}\right]^{-1}
\label{eq:Green_function}.
\end{equation}

\section{Model for an Andreev bound state}

In Eq.~(8) of the main text we introduce the general form of a tunneling Hamiltonian describing a normal lead coupled to a zero-energy Andreev bound state under the assumption of $s^z$ conservation. For concreteness, we shall now derive this Hamiltonian starting from a model of a single-level quantum dot coupled to a superconductor and to a normal lead.
The superconductor degrees of freedom can be integrated out, resulting in an effective low-energy Hamiltonian
\begin{equation}
\begin{array}{cc}
\multicolumn{2}{c}{ H= H_L + H_D + H_T }\\
H_L=\displaystyle\sum_{k s}\epsilon_k\psi_{k s}^\dag\psi^{\phantom \dag}_{k s} , & H_T = \displaystyle\sum_{k s}w_s\psi_{k s}^\dag d_s +{\rm h.c.}\\
\multicolumn{2}{c}{
H_D = \displaystyle\sum_{ss'} (\epsilon_0\delta_{ss'}+B\sigma^z_{ss'}) d^\dag_s d_{s'} + (\bar\Delta d^\dag_\uparrow d^\dag_\downarrow +{\rm h.c.}) ,}
\end{array}
\end{equation}
where $d^\dag_s$ creates a spin-$s$ electron in the dot, $\epsilon_0$ is the energy of the quantum dot level, $B$ is the Zeeman field, and $\bar\Delta$ is the induced pair potential in the dot. We assume that the charging energy is much smaller than $\bar\Delta$ and is therefore neglected.
Diagonalizing $H_D$, one has (up to a constant)
\begin{equation}
H_D = (\sqrt{\epsilon_0^2+\bar\Delta^2}-B)a^\dag a +(\sqrt{\epsilon_0^2+\bar\Delta^2}+B) b^\dag b ,
\end{equation}
where $a=\sin(\alpha) d^\dag_\uparrow-\cos(\alpha) d_\downarrow$, $b=\cos(\alpha) d_\uparrow+\sin(\alpha) d^\dag_\downarrow$, and $\cos(2\alpha)=\epsilon_0/\sqrt{\epsilon_0^2+\bar\Delta^2}$. To have a single Andreev bound state at zero energy we can now tune the Zeeman field to be $B=\sqrt{\epsilon_0^2+\bar\Delta^2}$. Finally, projecting $H_T$ onto the low-energy subspace described by $a$ and $a^\dag$ results in
\begin{equation}
H_T \simeq a^\dag\displaystyle\sum_{k}\left(w^\ast_\uparrow\sin(\alpha)\psi^{\phantom \dag}_{k \uparrow}+ w_\downarrow\cos(\alpha)\psi^\dag_{k\downarrow}\right) + {\rm h.c.} ,
\end{equation}
which is of exactly the same form as Eq.~(8) of the main text with $\tilde{t}_\uparrow=w^\ast_\uparrow\sin(\alpha)$ and $\tilde{t}_\downarrow=w_\downarrow\cos(\alpha)$.

\section{Spin rotations}
\label{spin_rot}
The value of the spin-resolved current correlation $P_{\uparrow\downarrow}$ obviously depends on the choice of spin basis, namely the axis along which spin-$\uparrow$ and spin-$\downarrow$ are defined. Given $P_{\uparrow\downarrow}$ in a particular spin-basis one can obtain $\tilde{P}_{\uparrow\downarrow}$, defined according to a rotated axis, by performing a unitary transformation on the scattering matrix and reinvoking Eq.~(5) of the main text.

Since the scattering matrix involves both electrons and holes degrees of freedom the rotating transformation is given by
\begin{equation}
\begin{array}{ccc}
\tilde{S}(E) = U^\dag S(E)U &;& U =\begin{pmatrix} e^{-i\frac{\theta}{2}\hat{\mathbf{n}}\cdot\boldsymbol{\sigma}}&\mathbf{0}\\ \mathbf{0}& e^{i\frac{\theta}{2}\hat{\mathbf{n}}\cdot\boldsymbol{\sigma}^\ast}
\end{pmatrix}
\end{array}
\end{equation}
where $\theta$ is the angle of rotation and $\hat{\mathbf{n}}$ is the rotation axis.

\begin{figure}
\includegraphics[clip=true, trim =3.3cm 9.3cm 1cm 10cm,scale=0.45]{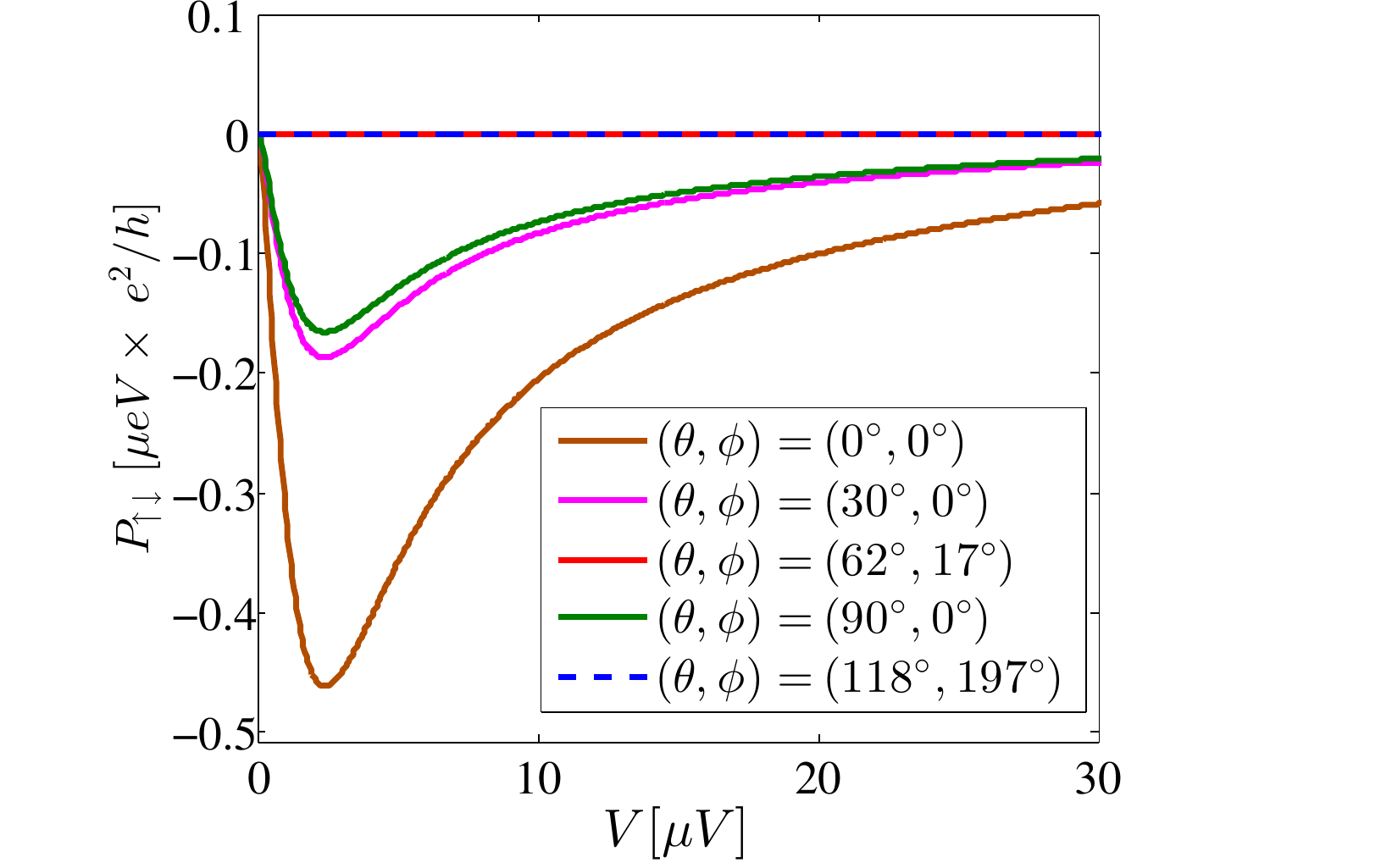}
\caption{Spin-resolved current correlations $P_{\uparrow\downarrow}$ as a function of bias voltage $V$ for the case of a Majorana bound state, at $T=0$. The spin-resolved currents are defined with respect to an axis which is given by the angles $\theta$ and $\phi$. By varying both $\theta$ and $\phi$ one arrives at a spin axis in which the spin-resolved current is perfectly polarized. One observes that  $P_{\uparrow\downarrow}$ vanishes for $(\theta,\phi)=(62^\circ,17^\circ)$ and for $(\theta,\phi)=(118^\circ,197^\circ)$. These are the axes for which $I_\downarrow\to0$ and $I_\uparrow\to0$ respectively.}\label{fig:spin_rotation}
\end{figure}

It was pointed out in the main text that for the case of a MBS one can always find a spin axis in which the current is completely spin polarized~\cite{He2014selective}. In that particular axis one should find that $P_{\uparrow\downarrow}\to0$. This property can be demonstrated for the system studied in this work (cf. Eq.~(12) of the main text). In Fig.~$3$a of the main text $P_{\uparrow\downarrow}$ was presented for a spin-axis which is rotated in the $xz$ plane. To witness perfect spin polarization of the spin-resolved current, however, one must rotate the spin-axis also along the azimuthal angle. In Fig.~\ref{fig:spin_rotation} we present $P_{\uparrow\downarrow}$ in the case of a MBS ($B=350\mu eV$) for a spin axis given by the angles $\theta$ and $\phi$, defined in the usual way.

\section{Finite-size effects}
\label{finite-size}

As mentioned in the main text, an isolated MBS gives rise to a negative spin-resolved current correlation $P_{\uparrow\downarrow}$. In a long but finite wire there can be a small overlap of the wave-functions of the MBSs at the two ends of the wire. It is instructive to examine the effect of this overlap on the spin-resolved current correlations.

In Fig.~\ref{fig:finite-size_plot} we present $P_{\uparrow\downarrow}$ as a function of bias voltage $V$ for different lengths of the superconducting section $L_{\rm S}$. The system parameters are otherwise the same as those in Fig.~$2$c of the main text. Due to the finite overlap of the Majorana end states there is a small region at very low voltages where $P_{\uparrow\downarrow}$ becomes positive. As $L_{\rm S}$ increases, and the overlap between the MBSs decreases, the positive-$P_{\uparrow\downarrow}$ region becomes shorter and its maximum value becomes smaller.

This finite-size effect is related to the one described in Fig.~$3$b of the main text. There we vary the overlap between the MBSs by changing the coherence length (increasing $B$) until reaching the limit where the overlap is maximal. Here, on the other hand, we vary the overlap by elongating the wire until reaching the limit where the overlap vanishes. Notice that in Fig.~$3$b of the main text we concentrate on values of $B$ for which a zero-energy states is present inspite of the spatial overlap of the MBSs (cf. Fig.~$2$a of the main text). Here, on the other hand, the overlap is accompanied by an energy splitting of the MBSs.

\section{Supra-gap voltages}
\label{ultra_gap_voltage}

\begin{figure}
  \subfloat[]{\includegraphics[clip=true, trim =4.6cm 9.2cm 4.05cm 9.9cm,width=0.24\textwidth]{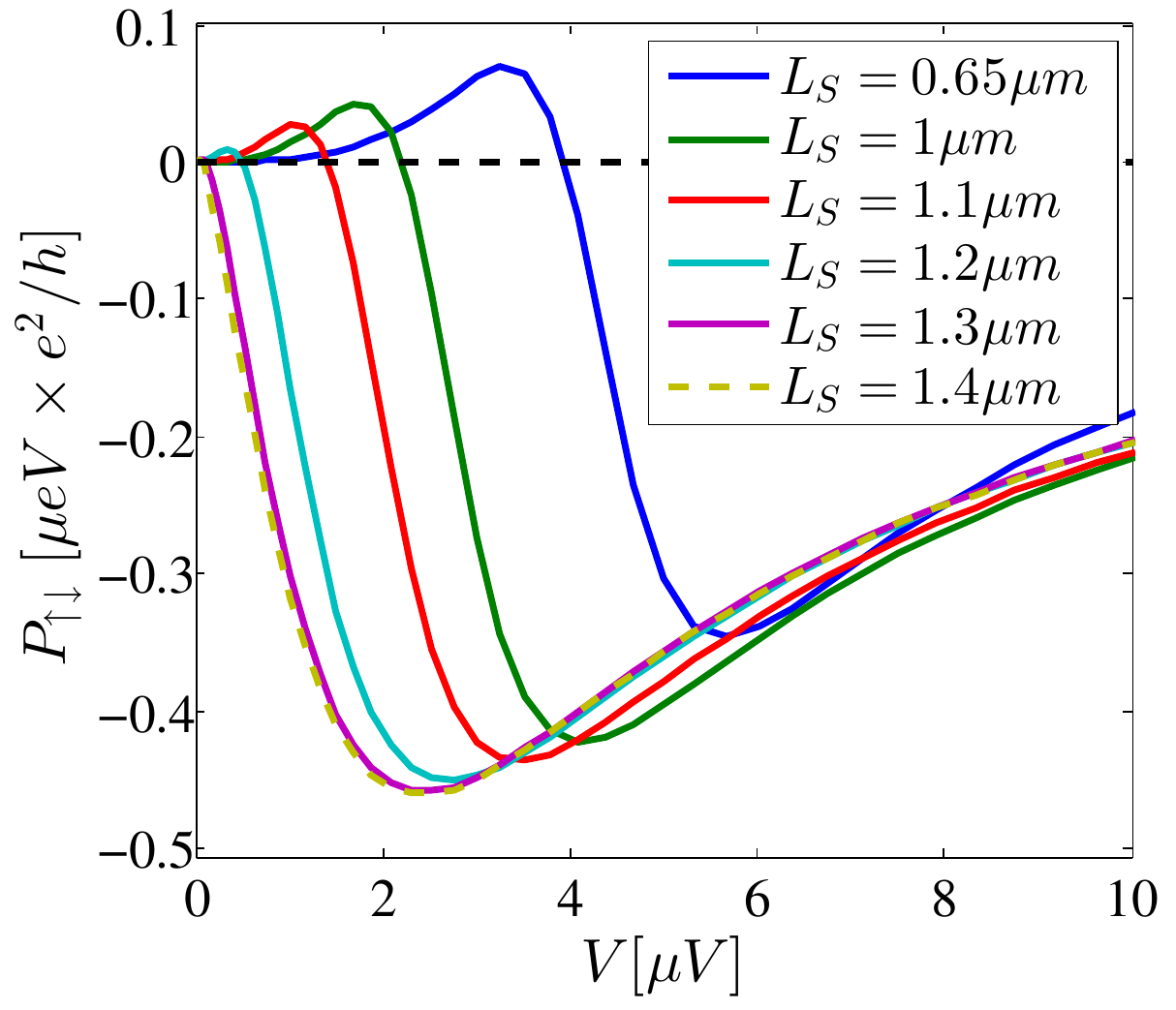}\label{fig:finite-size_plot}}
    \subfloat[]{\includegraphics[clip=true, trim =4.6cm 9.2cm 4.05cm 9.9cm,width=0.24\textwidth]{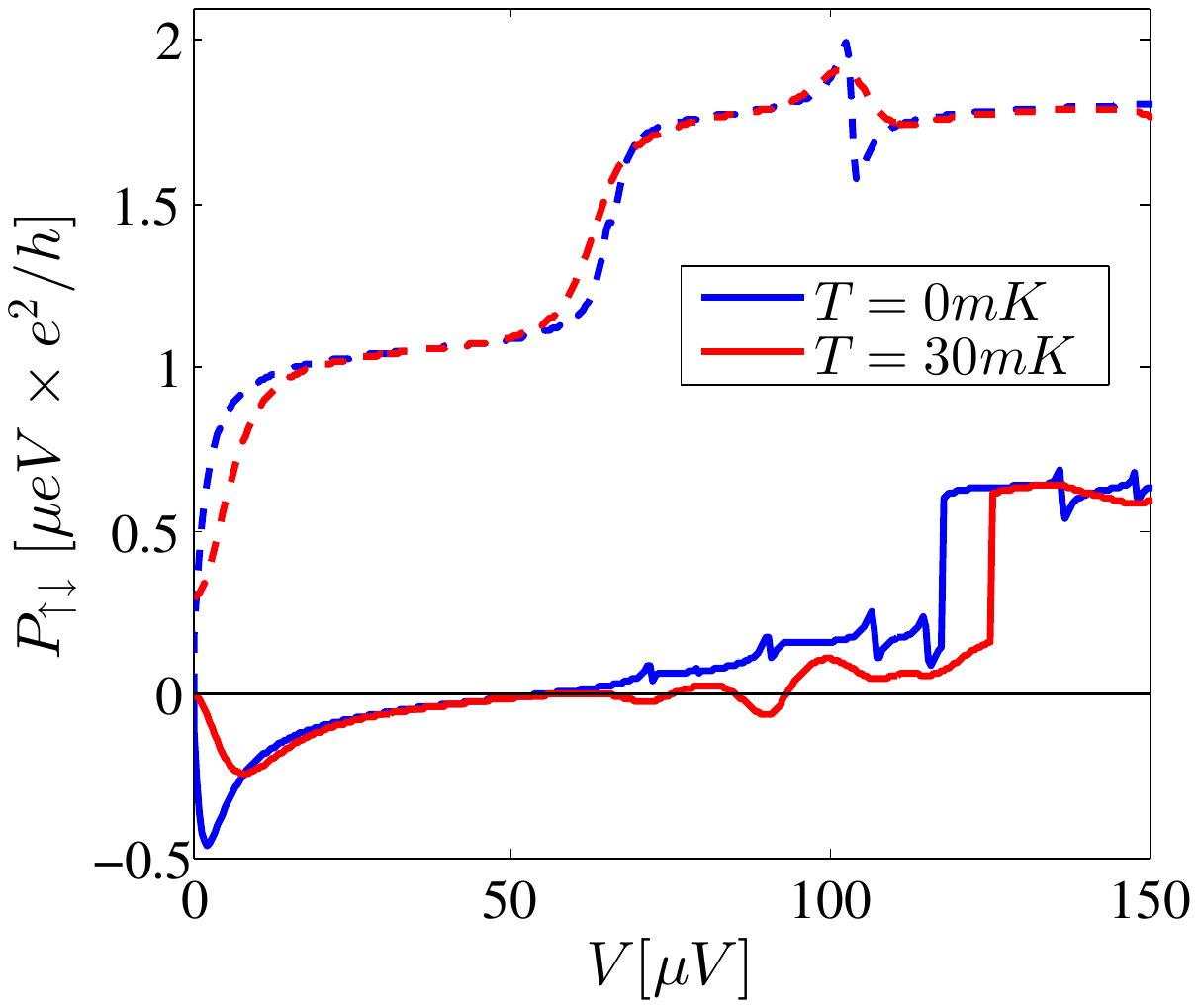}\label{fig:ultra_gap_voltage}}
\caption{(a) Spin-resolved current correlation $P_{\uparrow\downarrow}$ vs bias voltage $V$ at $T=0$ for different lengths of the superconducting section $L_{\rm S}$ for the case of a Majorana bound state. The region of positive $P_{\uparrow\downarrow}$ at small voltages is due to the overlap between the two majorana fermions at the wire ends. As $L_{\rm S}$ increases the overlap becomes smaller. As a result the positive region becomes shorter and its maximum value becomes smaller. (b) $P_{\uparrow\downarrow}$ vs $V$ at zero and finite temperatures for a Majorana bound state (solid lines) and for an Andreev bound state (dashed lines). At voltages higher than the gap (which is about $50\mu eV$) the behavior is non universal as $P_{\uparrow\downarrow}$ picks up contributions from higher excited states. System parameters are the same as in Fig.~$2$ of the main text.}
\end{figure}

In the main text we have presented results for voltages which are smaller than the excitation gap in the system, which is about $50\mu eV$ (cf. Fig.~$2$c and Fig.~$2$d of the main text). As was mentioned, at higher voltages the features of $P_{\uparrow\downarrow}$ are no longer universal as $P_{\uparrow\downarrow}$ picks up contributions from higher-energy resonances. For completeness we  present in Fig.~\ref{fig:ultra_gap_voltage} the spin-resolved correlation $P_{\uparrow\downarrow}$ for higher bias voltage for both the MBS case ($B=350\mu eV$) and the ABS case ($B=90\mu eV$). All parameters are the same as in Fig.~$2$ of the main text.

\section{Additional results}
\label{addit_res}

In the main text we have presented results (cf. Fig.~$2$ and Fig.~$3$) of a numerical simulation of a system having the same parameters as those of a recent experiment by Mourik {\it et al.}~\cite{mourik2012signatures}. The conclusions and the main features of our study are general and do not depend on the specific choice of parameters.

We have repeated the calculation for a system having parameters similar to those of an experiment by Das {\it et al.} ~\cite{Das2012zero}, namely $E_{\rm so}=m_e\alpha_R^2/2=70\mu eV$, $\Delta_0=45\mu eV$, and $l_{\rm so}=1/(m_e\alpha_R)=130nm$. We take the length of the wire to be $L=2.6\mu m$ with $L_{\rm S}=1.3\mu m$. For chemical potential $\mu=0$, the system is in its topological phase when $B>B_c=45\mu eV$, with two MBSs residing at each end of the wire.

\begin{figure}
    \subfloat[]{\includegraphics[clip=true, trim =4.6cm 9.25cm 4.15cm 9.9cm,width=0.24\textwidth]{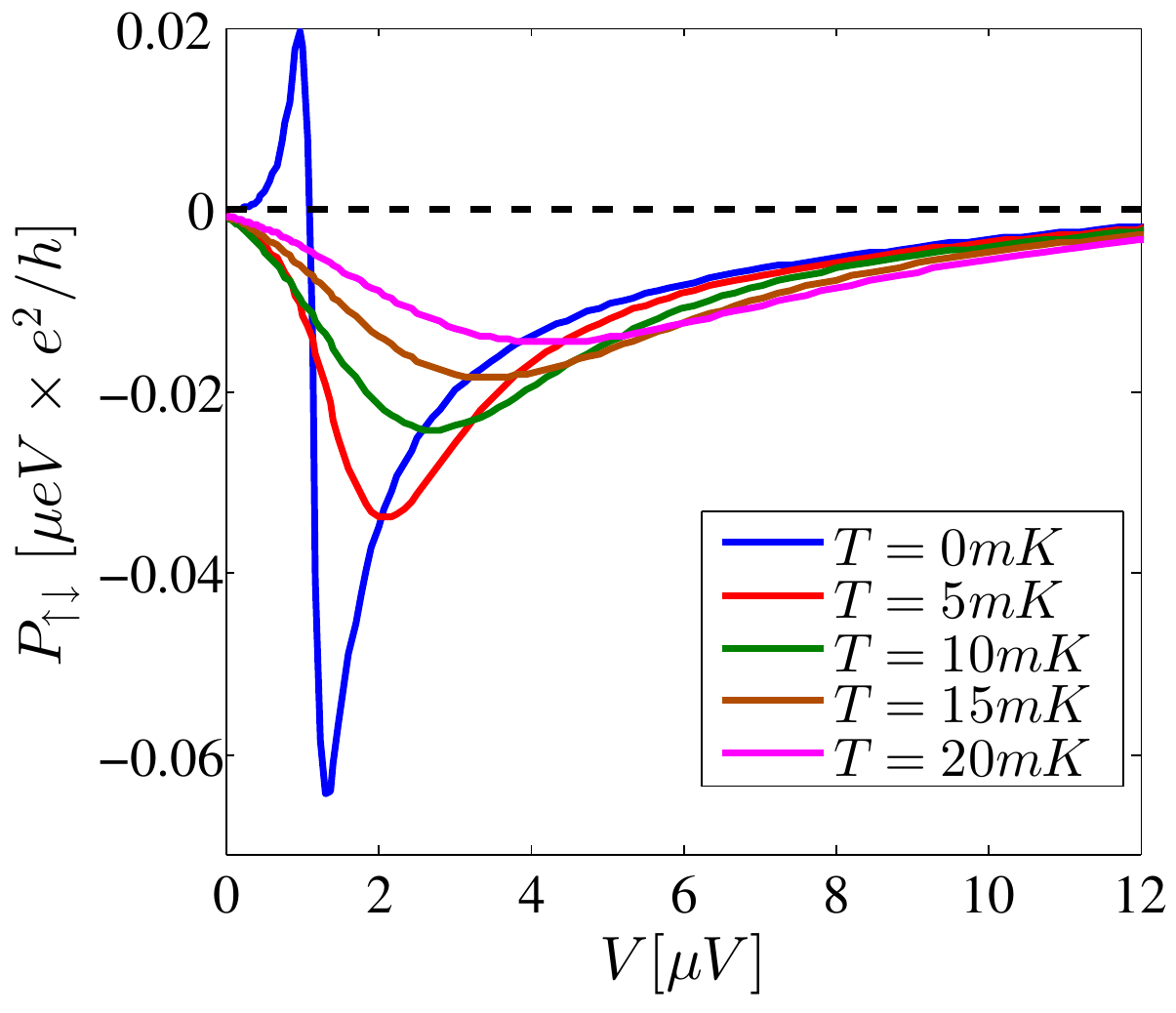}\label{fig:Heiblum_MBS}}
    \subfloat[]{\includegraphics[clip=true, trim =4.6cm 9.25cm 4.15cm 9.9cm,width=0.24\textwidth]{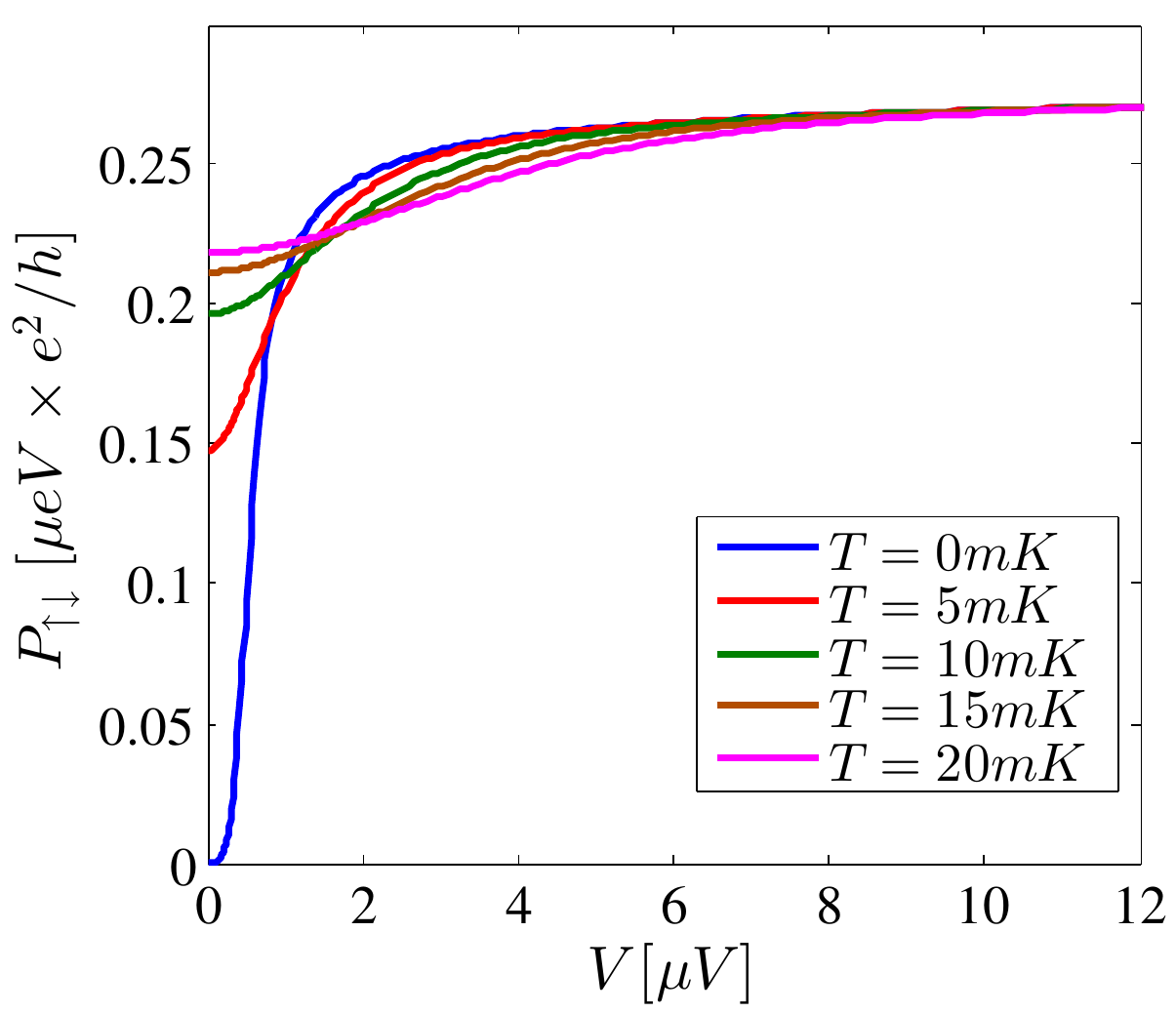}\label{fig:Heiblum_ABS}}

        \subfloat[]{\includegraphics[clip=true, trim =3.8cm 9.25cm 4.15cm 10cm,width=0.24\textwidth]{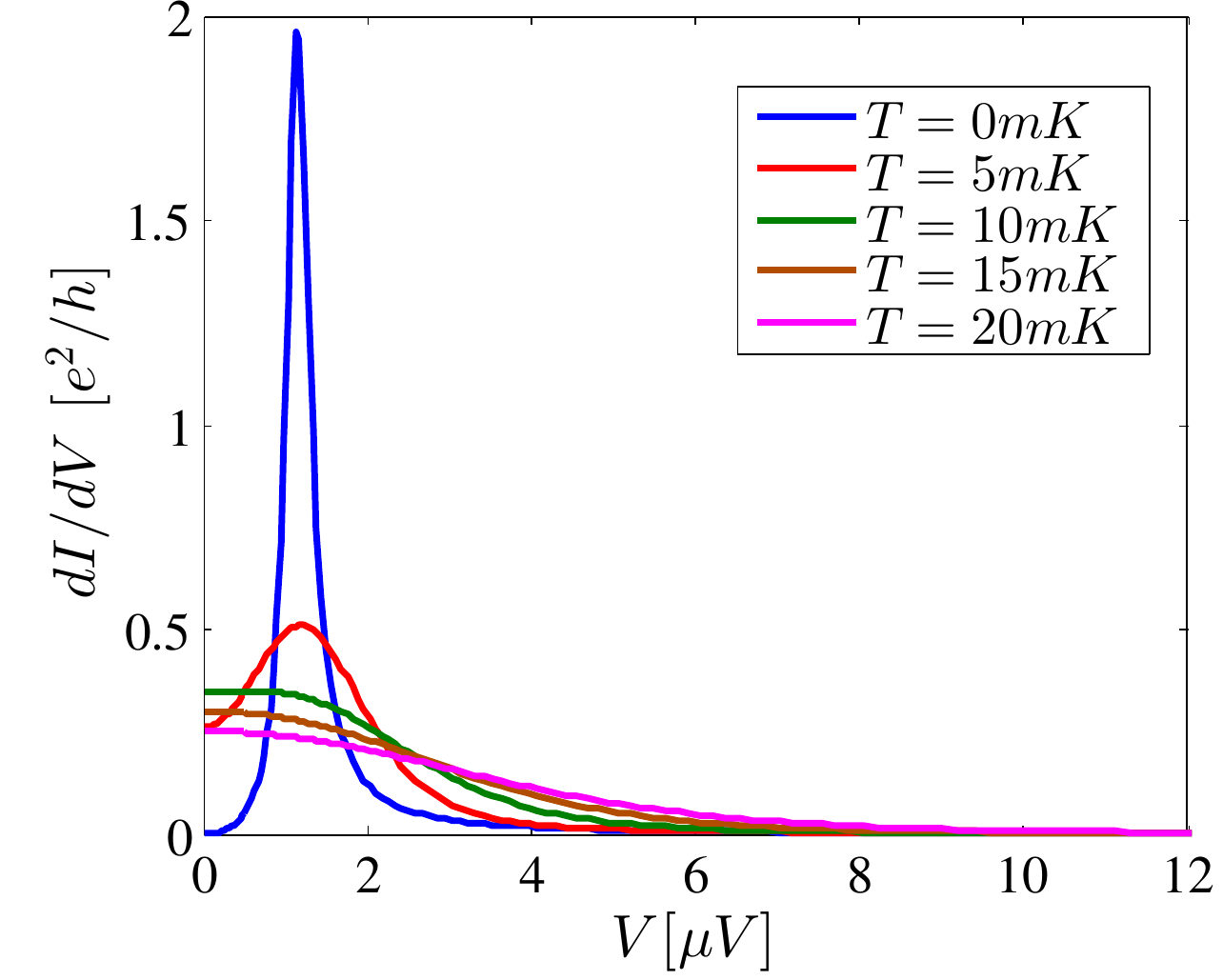}\label{fig:Heiblum_G_MBS}}
    \subfloat[]{\includegraphics[clip=true, trim =3.6cm 9.25cm 4.15cm 10cm,width=0.24\textwidth]{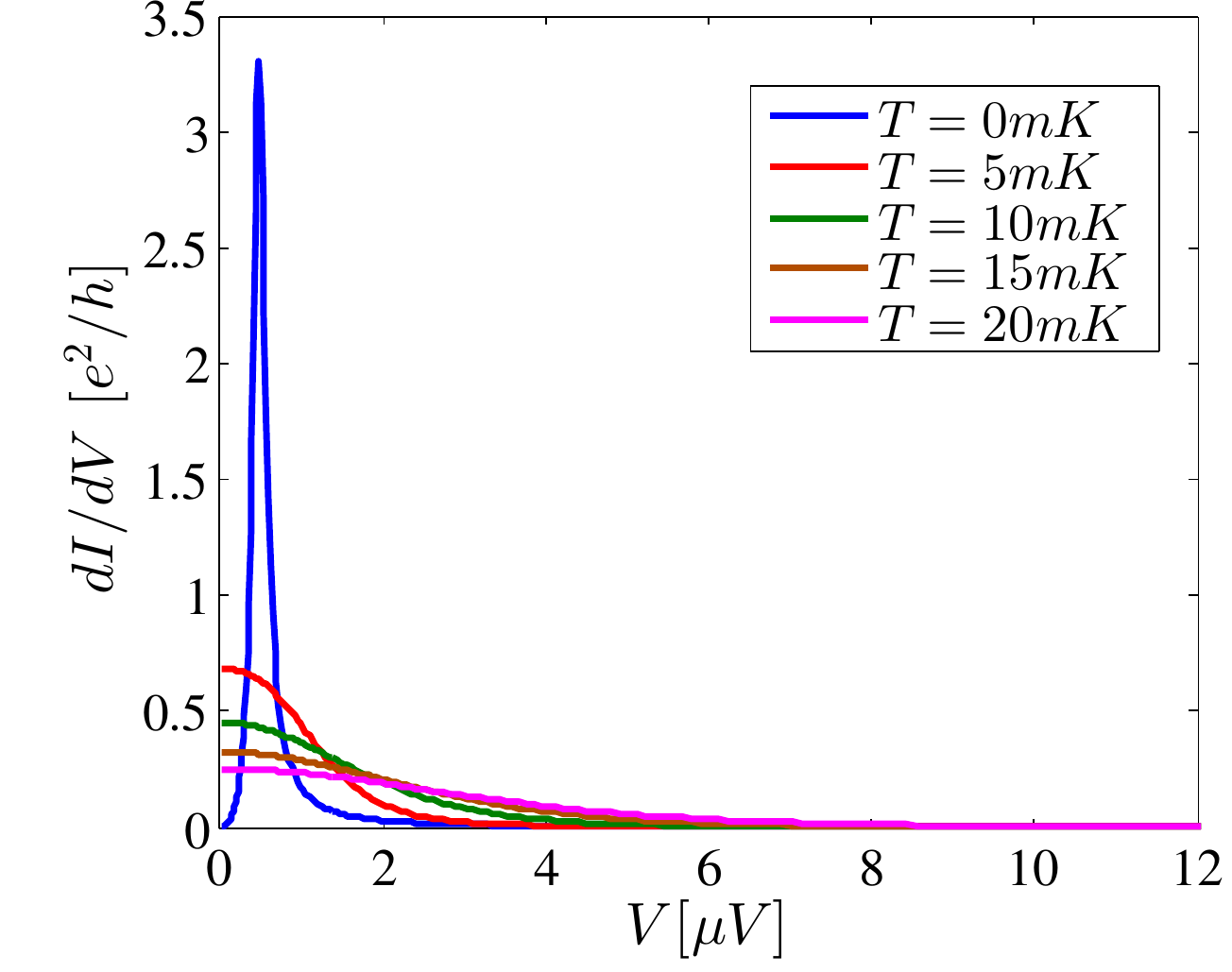}\label{fig:Heiblum_G_ABS}}
\caption{(a)-(b) Spin-resolved current correlations $P_{\uparrow\downarrow}$ for (a) a Majorana bound state and for (b) an Andreev bound state. (c)-(d) Differential conductance $dI/dV$ for (c) a Majorana bound state and for (d) an Andreev bound state. The system parameters are similar to those of a recent experiment by Das {\it et al.}~\cite{Das2012zero}. We take the length of the wire to be $L=2.6\mu m$ with $L_{\rm S}=1.3\mu m$, and $\mu=0\mu eV$.}\label{fig:Heiblum_params}
\end{figure}

In Fig.~\ref{fig:Heiblum_params} we present results for $P_{\uparrow\downarrow}$ and $dI/dV$ for the case of a MBS (Zeeman field of $B=68\mu eV>B_c$, pointing in the $x$ direction), and for a case of an accidental ABS (Zeeman field $B=16\mu eV<B_c$, pointing in the $z$ direction). While the differential conductance spectra for the ABS and for the MBS look similar, the behavior of the spin-resolved correlations is very distinct. For the MBS $P_{\uparrow\downarrow}$ is mostly negative and approaches zero at high bias voltages. The region of positive $P_{\uparrow\downarrow}$ at very small voltages is due to finite size effects as explained above. In the case of the ABS, on the other hand, $P_{\uparrow\downarrow}$ is positive and saturating at a constant nonzero value.

\end{document}